\documentclass[preprint,11pt]{elsarticle}
\usepackage[margin=1in]{geometry}
\usepackage{comment}
\usepackage{amssymb}
\usepackage{amsmath}
\newcommand{\ds}              { \displaystyle }
\newcommand{\bfm}[1]             { \mathbf{#1}     }
\newcommand{\uu}              { \bfm{u} }
\newcommand{\xx}              { \bfm{x} }
\newcommand{\ww} {\bfm{w}}
\newcommand{\FF}              { \bfm{F} }

\newcommand\dt[1]{\frac{\partial #1}{\partial t}}

\newcommand\dx[1]{\frac{\partial #1}{\partial x}}
\newcommand\dy[1]{\frac{\partial #1}{\partial y}}
\newcommand\area[2]{\left(#1,#2\right)_{\Omega_e}}
\newcommand\edge[2]{\langle #1, #2 \rangle_{\partial \Omega_e}}
\newcommand{\PD}{\text{M}\Pi_h}
\newcommand{\PZ}{\text{M}\Pi^d_h}
\newcommand{\PU}{\text{M}\Pi^f_h}

\usepackage{lineno}
\usepackage{graphicx}
\usepackage{subfigure}
\usepackage[colorlinks]{hyperref}
\AtBeginDocument{%
  \hypersetup{%
    linkcolor=blue,%
    citecolor=red,%
  }%
}%
\usepackage{booktabs}
\usepackage{float}
\usepackage{tikz}
\usepackage{tabularx}
\usetikzlibrary{shapes.geometric, arrows}
\tikzstyle{process} = [rectangle, minimum width=3cm, minimum height=1cm, text centered, draw=black] %fill=orange!30]

\begin{document}

\begin{frontmatter}


 \title{A Discontinuous Galerkin Finite Element Model for Compound Flood Simulations}
 \author[inst1]{Chayanon Wichitrnithed\corref{cor1}}
 \ead{namo@utexas.edu}
 \cortext[cor1]{Corresponding author}

\affiliation[inst1]{organization={The Oden Institute for Computational Engineering and Sciences, The University of Texas at Austin},%Department and Organization
            addressline={201 E. 24th St. Stop C0200},
            city={Austin},
            postcode={78712},
            state={Texas},
            country={United States of America}}
\affiliation[inst2]{organization={The Department of Data Science, The Norwegian University of Life Science},%Department and Organization
            addressline={Drøbakveien 31},
            city={Ås},
            postcode={1433},
            country={Norway}}
\affiliation[inst3]{organization={Department of Scientific Computing and Numerical Analysis, Simula Research Laboratory},%Department and Organization
            addressline={Kristian Augusts gate 23},
            city={Oslo},
            postcode={0164},
            country={Norway}}

\author[inst1,inst2,inst3]{Eirik Valseth}
\author[inst4]{Ethan J.\ Kubatko}
\author[inst4]{Younghun Kang}
\author[inst4]{Mackenzie Hudson}
\author[inst1]{Clint Dawson}

\affiliation[inst4]{organization={The Department of Civil, Environmental, and Geodetic Engineering, The Ohio State University},%Department and Organization
            addressline={2070 Neil Ave},
            city={Columbus},
            postcode={43210},
            state={Ohio},
            country={United States of America}}

\begin{abstract}
Many tropical cyclones, e.g., Hurricane Harvey (2017), have lead to significant rainfall
and resulting runoff. When the runoff interacts with storm
surge, the resulting floods can be greatly amplified and lead to effects that cannot be correctly
modeled by simple superposition of its distinctive sources. In an effort to develop accurate numerical simulations of runoff, surge, and compounding floods, we develop a local discontinuous Galerkin method for modified shallow water equations. In this modification, nonzero sources to the continuity equation are included to incorporate rainfall into the model using parametric rainfall models from literature as well as hindcast data. The discontinuous Galerkin spatial discretization is accompanied with a strong stability preserving explicit Runge Kutta time integrator. Hence, temporal stability is ensured through the Courant-Friedrichs-Lewy (CFL) condition and we exploit the embarrassingly parallel nature of the developed method using  MPI parallelization.

We demonstrate the capabilities of the developed method though a sequence of physically relevant numerical tests, including small scale test cases based on laboratory measurements and large scale experiments with Hurricane Harvey in the Gulf of Mexico. The results highlight the conservation properties and robustness of the developed method and show the potential of compound flood modeling using our approach.
\end{abstract}




\end{frontmatter}

\clearpage

\section{Introduction} \label{sec:introduction}

In this paper, we present recent advances and developments of discontinuous Galerkin (DG) finite element (FE) methods in shallow water flows in coastal, riverine, and overland regions. In particular, we focus on the application of DG methods to the shallow water equations (SWE) for  flood events in which hurricane storm surge interacts with one or more other sources of water, i.e., compound flood events~\cite{wahl2015increasing}.
Modeling these events in coastal regions remains a significant challenge as the number of physical processes that play a role are significant, e.g., storm surge, river discharge, tides, rainfall, and winds.
 These processes typically interact in a highly nonlinear fashion and require careful treatment to ensure accurate modeling~\cite{loveland2021developing,santiago2019comprehensive,orton2020flood,kumbier2018investigating}. Our approach to this modeling challenge is the development of a comprehensive DG solver which computes FE approximations to modified SWE with nonzero source terms in a single domain covering ocean, rivers, and overland areas.  Hence, in this work, we focus on compound flooding from the following sources: storm surge, river runoff, and rainfall.

Flooding from each of the sources has been extensively studied, and there are numerous models in existence.
\begin{comment}
Here, we do not aim to provide a complete list here but mention some well-established and commonly used models. For storm surge, the ADvanced CIRCulation (ADCIRC) model~\cite{Pringle2020,luettich1992adcirc}, the Sea, Lake, and Overland Surges from Hurricanes (SLOSH)~\cite{jelesnianski1992slosh}, and Delft3D~\cite{veeramony2017forecasting} have all been extensively validated for past hurricane events. In addition, there are also a significant number of validated models for riverine and rain-induced overland flows, such as the Hydrologic Engineering Center River Analysis System (HEC-RAS) model~\cite{brunner2021ceiwr,brunner1995hec} and the Gridded Surface Hydrological Analysis  (GSSHA) model~\cite{downer2004gssha}. These two sets of models are typically applied over vastly different scales, from watershed scale to global ocean scale.
\end{comment}
A significant portion of current modeling efforts of compound flooding are based on coupling existing models for individual physical processes, see~\cite{santiago2019comprehensive} for a comprehensive review. The methodologies for coupling such models can vary in complexity from simple one-way coupling in which input from one model is the output from another, to   fully coupled multi-physics models in which all governing physics are resolved simultaneously. The models of lower complexity typically consist of well established, validated models for each individual flood component. However, recent work utilizing the aforementioned surge models, e.g. \cite{loveland2021developing} has shown the applicability of ADCIRC in  modeling interactions between riverine flows and storm surge through careful construction of the FE mesh and model inputs.

To model compound flooding, we rely on the SWE as the governing model. The SWE are a set of nonlinear transient partial differential equations (PDEs), and thus, analytic solutions cannot be established for cases beyond that of academic interest. Hence, accurate numerical approximation is critical. Current (physics-based) modeling efforts based on numerical approximation of transient PDEs such as the SWE can broadly be categorized to be of FE, finite volume (FV), and finite difference (FD) approximations for the spatial differential operators. In general, the temporal differential operators are discretized using implicit, explicit, or a combination of implicit and explicit FD schemes. Space-time approaches in which both spatial and temporal operators are discretized using FE methods also exist, see, e.g. \cite{takase2010space}. However, due to the computational cost of such approaches these are generally not used in large-scale shallow water modeling. In our approach to model compound flooding from surge, runoff, and rainfall, we employ the  previous work on explicit-in-time DG methods for SWE described in \cite{kubatko2006hp} and \cite{dawson2011discontinuous}.
DG methods are chosen as they possess several features making them very well suited to solve the SWE, including: local conservation, unstructured meshes, $p-$adaptivity, ease of parallelization, etc.~\cite{cockburn1998runge}. The main trade-off is the higher degrees of freedom which increases execution time when compared to a continuous Galerkin formulation.

%\eirik{Probably a paragraph to review the parametric rainfall models here}

One critical addition to the above DG model in this work is rainfall in the form of a source term appearing in the depth-averaged continuity equation of the SWE (see Equation \eqref{eq:SWE} of the next section). Two methods of approximating the rainfall term are explored here. One approach is the direct interpolation of rainfall rates onto the mesh from observed rainfall data, which are available from a number of sources at various spatial and temporal resolutions (see \cite{NCAR:Datasets} for a comprehensive overview of precipitation data sets). A second approach, applicable during tropical cyclones (TCs), is the implementation of a parametric rainfall model directly within the DG model. This approach takes advantage of the fact that, during TCs, rainfall patterns in the vicinity of the storm exhibit defined structures that can be exploited by a parametric model, which uses simple analytic expressions based on a small number of (predicted or observed) storm parameters that are made publicly available by the National Hurricane Center pre and post storm events.
This allows us to directly compute rain intensity as a source term solely using  existing wind input data.
%This type of approach is already implemented in our DG model for constructing wind fields during TCs and provides an existing code infrastructure that can be leveraged for computing rainfall rates.

The use of parametric TC rainfall models within storm surge models has only recently been explored in \cite{Bilskie:2021}, wherein a parametric TC rainfall model was implemented within ADCIRC and evaluated on a set of twelve synthetic (not historic) storms in terms of the increase to peak water levels it produced versus modeling surge alone.
We also note the work in \cite{Dresback:2019}, wherein a coupled modeling system, called STORM-CoRe, was developed that makes use of a parametric TC rainfall model to provide input to a distributed hydrologic model, which, in turn, provides stream flows to ADCIRC via inflow boundary conditions.
As noted in their later work \cite{Dresback:2023}, while this coupled approach accounts for rainfall effects in the ``upland areas,'' it does not account for it in the coastal regions.
To rectify this, similar to the approach taken by \cite{Bilskie:2021} noted above, they add a rainfall source term directly within ADCIRC.
However, the specific use of a parametric TC rainfall model to provide the rainfall source term is not investigated; instead, rainfall estimates are provided by Multi-Radar/Multi-Sensor (MRMS) data \cite{Zhang:2018}.

In the following, we introduce, verify, and validate our compound flood modeling methodology. In Section \ref{sec:DG_discr}, we introduce the governing model PDE, modified to account for rainfall. This section also introduce the DG discretization of the SWE and the parametric rainfall models used to generate rainfall source data.
In addition, we provide an overview of the implementation details as well as other recent additions to the numerical model we employ.
Next, in Section \ref{sec:experiments}, we present a comprehensive verification and validation of our methodology including benchmark tests of conservation properties as well as large scale hurricanes with compound flood effects. Finally, in Section~\ref{sec:conclusions}, we draw conclusions and discuss potential future research directions.

\section{Methods} \label{sec:DG_discr}
\subsection{Governing Equations}
\begin{figure}[h!]
    \centering
    \scalebox{.625}{\input{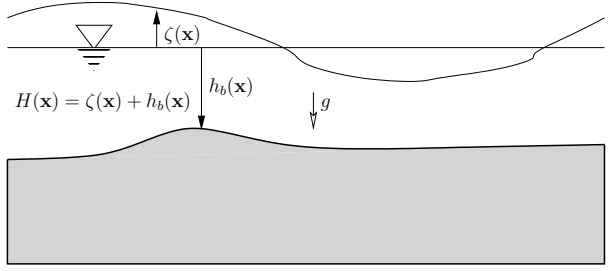} }%
    \caption{Definition of shallow water elevations. The horizontal line denotes the geoid, where $\zeta = h_b = 0$.}
    \label{fig:elevation_def}
\end{figure}

The governing model we use for shallow water flow are the two-dimensional SWE which consist of the conservative depth-averaged equations of mass conservation as well as $x$ and $y$ momentum conservation \cite{tan1992shallow}:
\begin{equation} \label{eq:SWE}
\begin{array}{ll}
\text{Find }  (\zeta, \uu)   \text{ such that:}  \qquad \qquad    \\ \\
\ds \frac{\partial  \zeta}{\partial t} + \nabla \cdot (H{\uu})  = R, \text{ in } \Omega, & \\ \\
\ds\frac{\partial (Hu_x)}{\partial t} + \nabla \cdot \left( Hu_x^2 + \frac{g}{2}(H^2-h_b^2), Hu_xu_y \right) - g\zeta \frac{\partial h_b}{\partial x} + \tau_b Hu_x = F_x, \text{ in } \Omega, & \\ \\
\ds\frac{\partial (Hu_y)}{\partial t} + \nabla \cdot \left( Hu_xu_y, Hu_y^2 + \frac{g}{2}(H^2-h_b^2) \right) - g\zeta \frac{\partial h_b}{\partial y} + \tau_b Hu_y = F_y, \text{ in } \Omega,
 \end{array}
\end{equation}
where $\zeta$ is the free surface elevation (positive upwards from the geoid),  $h_b$ the bathymetry (positive downwards from the geoid), $H$ is the total water column (see Figure \ref{fig:elevation_def}),  $\uu = \{ u_x,u_y\}^{\text{T}}$ is the depth-averaged velocity field, $\tau_b$ is the bottom friction factor, $R$ is the rainfall rate, and the source terms $F_x,F_y$ represent potential relevant sources which induce flow, including, e.g., Coriolis force, tidal potential forces, wind stresses, and wave radiation stresses. $\Omega$ is the computational domain, e.g., the coastal ocean, and its boundary $\Gamma$ is identified by three distinctive sections $\Gamma = \Gamma_{ocean}\cup\Gamma_{land}\cup\Gamma_{river}$.

The PDEs \eqref{eq:SWE} differs from the previous works \cite{kubatko2006hp,dawson2011discontinuous} in the inclusion of the (potentially nonzero) right hand side of the mass conservation equation, the rainfall rate $R$. To complete an initial boundary value problem (IBVP), \eqref{eq:SWE} is augmented with proper  initial and boundary conditions. In addition to the open ocean (specified elevation) and land boundary condition (zero normal flow) used in \cite{dawson2011discontinuous}, we have river flow (specified normal flow) boundaries:
\begin{equation} \label{eq:river_bc}
\begin{array}{ll}
\ds \uu \cdot \bfm{n} = q_{river} \text{ on } \Gamma_{river},
 \end{array}
\end{equation}
where $q_{river}$ are the specified river flows of unit meters squared per second.

\subsection{Discontinuous Galerkin Formulation}
The DG formulation and solution method used within our framework is based on the previous works \cite{kubatko2006hp,dawson2011discontinuous}. Hence, we do not include a comprehensive introduction to the details of the formulation but rather a brief overview to make this presentation sufficiently self contained.
By defining the vector of unknowns $\mathbf{w} = \{\zeta, Hu_x, Hu_y \}^\text{T}$, the source vector $\mathbf{s}$ and the flux matrix $\mathbf{F}(\ww)$ as:
\begin{equation*}
    \mathbf{w} =
                 \begin{bmatrix}
                 \zeta, & uH, & vH
                 \end{bmatrix}^T \\
\end{equation*}
\begin{equation*}
    \mathbf{s} =
                 \begin{bmatrix}
                   R, & g\zeta\dx{h_b} + F_x,  & g\zeta\dy{h_b} + F_y
                 \end{bmatrix}^T \\
\end{equation*}
\begin{equation*}
    \mathbf{F}(\ww) =
                 \begin{bmatrix}
                   uH & vH \\
                   Hu^2 + g(H^2-h_b^2) & Huv \\
                   Huv & Hv^2 + g(H^2-h_b^2)
                 \end{bmatrix}
\end{equation*}
we can write the three equations in the form
\begin{align}
    \dt{\mathbf{w}} + \nabla \cdot \bfm{F}(\ww) = \mathbf{s}, \quad i = 1,2,3.
\end{align}
The corresponding weak formulation is obtained by multiplying each equation by a test function $\mathbf{v}$ and integrating over each element $e$, and subsequent integration by parts:
\begin{align} \label{eq:elem_weak_form}
    \area{\dt{\mathbf{w}}}{\mathbf{v}} - \area{\nabla \mathbf{v}}{\bfm{F}} + \edge{\hat{\bfm{F}} \cdot \bfm{n}}{\mathbf{v}} = \area{\mathbf{s}}{\mathbf{v}}
\end{align}
where $\hat{\bfm{F}}$ represents a choice of numerical flux. The role of the numerical flux is to couple adjacent, discontinuous elements and correctly capture the flow characteristics across the edges; this is necessary for the solution to be stable \cite[Chapter~2]{Hesthaven_undated-bz}.
The solution $\ww$ is then spatially approximated as $\ww_h$ using an orthogonal Dubiner basis \cite{Dubiner1991-ts} $\phi_{ij}$ as:
\begin{align} \label{dubiner}
    \ww_h = \sum_i \sum_j \widetilde{\ww}_{ij} \phi_{ij}
\end{align}
where $\widetilde{\ww}_{ij}$ are the modal degrees of freedom. The indices $i,j$ indicate the order of the polynomial: $\phi_{ij}$ is of order $i+j$. This basis is chosen as it produces a diagonal mass matrix, and higher-order approximations can be obtained by simply adding more terms. Substituting this into the weak formulation above as well as an identical choice for the test function $\mathbf{v}$ reduces the problem to a system of ODEs in the form:
\begin{align} \label{eq:RKSSP}
    \frac{d}{dt} (\ww_h) = L_h (\ww_h),
\end{align}
where the terms from~\eqref{eq:elem_weak_form} are collected into $L_h$.
We then use an optimized version of the explicit Strong Stability Preserving Runge-Kutta scheme (SSPRK) based on Shu and Osher \cite{Shu1987-ws,Shu1988-eq} to discretize~\eqref{eq:RKSSP} in time and  compute the difference approximation $\ww_h^{n+1}$, at the next timestep. This scheme has been specifically optimized with respect to DG spatial discretizations to improve stability regions and therefore increase the possible time step size for a given order \cite{Kubatko2014-nc}.

\subsection{Implementation}

To implement the SSPRK DG-method of our SWE, we extend the Discontinuous Galerkin Shallow Water Equation Model (DG-SWEM) first developed as part of~\cite{kubatko2006hp}.
The code is parallelized through MPI and is compatible with meshes and inputs developed for ADCIRC.
In the implementation, $p-$order DG approximations as introduced in \eqref{dubiner} are available with several choices for the numerical flux $\hat{\bfm{F}}$ such as Roe \cite{Roe1981-ls} and Local Lax-Friedrichs (LLF) \cite{Leveque2004-bl}, an advanced wetting and drying algorithm~\cite{bunya2009wetting},  as well as an optional slope limiter that can be applied in cases where physical shocks arise.

%The basic procedure of the code is outlined in Figure~\ref{flowchart}.
The spatial discretization of the DG formulation is local to each element; that is, the approximation of the integral terms appearing in~\eqref{eq:RKSSP} can be performed element-by-element without having to solve a global mass matrix system as in the standard continuous Galerkin FE method.
Specifically, we first compute the edge integrals $\edge{\hat{\bfm{F}} \cdot \bfm{n}}{v}$ of each element and then the area integrals $\area{s}{v} + \area{\nabla v}{\bfm{F}}$ of each element.
%
%
%\begin{figure}[h!]
%\centering
%\tikzstyle{arrow} = [thick,->,>=stealth]
%\begin{tikzpicture}[node distance=2cm]
%\node (start) [process] {Start};
%\node (init) [process, below of=start] {Initialize variables and arrays};
%\node (edge1) [process, below of=init] {Compute boundary condition edge integrals};
%\node (edge1) [process, below of=init] {Compute elevation-specified edge integrals};
%\node (edge2) [process, below of=edge1] {Compute land boundary edge integrals};
%\node (edge3) [process, below of=edge2] {Compute inner edge integrals};
%\node (forcing) [process, below of=edge1] {Compute  forcing, including rain at each node};
%\node (area) [process, below of=forcing] {Compute the  integrals of each element};
%\node (next) [process, below of=area] {Compute the solution at the next timestep};
%\node (end) [process, below of=next] {Finish};

%\draw [arrow] (start) -- (init);
%\draw [arrow] (init) -- (edge1);
%\draw [arrow] (edge1) -- (edge2);
%\draw [arrow] (edge2) -- (edge3);
%\draw [arrow] (edge1) -- (forcing);
%\draw [arrow] (forcing) -- (area);
%\draw [arrow] (area) -- (next);
%\draw [arrow] (next) -- (end);
%\draw [arrow] (next) -|- (start);
%\draw [arrow] (next.east) -- ++ (2cm,0) |- node[above]{Time step loop} (edge1);
%\end{tikzpicture}

%\caption{Flowchart of DG-SWEM}
%\label{flowchart}
%\end{figure}
%
%

The entry point for rainfall is nodal, i.e. rain intensity $r$ will have to be specified at each node in the grid. For example, it could be spatially interpolated from historical data or estimated using \textit{parametric models} as discussed in the next section. To transform the nodal rain intensity to the elemental source term $R$ we simply compute the average rain intensity, i.e. in the term $\area{R}{v}$ we have
\begin{align} \label{eq:rain}
    R = \frac{R(n_e^1) + R(n_e^2) + R(n_e^3)}{3}
\end{align}
where $n_e^1,n_e^2,n_e^3$ are the nodes of element $e$.
%Note that this term is added for all elements, whether wet or dry. This allows dry elements to become eventually wet from the rain and not only from mass transfer from neighboring elements.

Apart from the addition of rainfall forcing into the DG-SWEM codebase, we have also added the capability of transient river boundary conditions as in equation ~\eqref{eq:river_bc}. In practice, these are prescribed by properly selecting the numerical flux $\hat{\bfm{F}}$ for the element edges that are river boundaries. Note that these do not have to be on the edges of the mesh, and internal river boundaries can also be applied in this fashion. The final point of which we have improved the capabilities of DG-SWEM is the incorporation of a new parametric hurricane wind and pressure field model called the  Generalized Asymmetric Holland Model (GAHM) \cite{Gao2018-ik}. This parametric model is based on the classical Holland Model \cite{holland1980analytic} and has been designed specifically to handle ADCIRC hurricane input. These parametric models use National Hurricane Center (NHC) forecasts (or hindcasts after the hurricanes have dissipated) which includes time series data such as storm track, central pressure, and  maximum winds.  We refer to the ADCIRC Wiki page\footnote{\url{https://wiki.adcirc.org/Generalized\_Asymmetric\_Holland\_Model\#cite\_note-1}} for further details on the GAHM. From this parametric hurricane wind and pressure model, we ascertain wind and pressure fields that are applied at all points in the computational domain.

\subsection{Wetting and drying} \label{sec:wetdry}
Since the SWE are only defined for wet regions ($H > 0$), we must essentially work with moving boundary problems. To avoid the complexity of modifying the mesh to reflect the changing boundaries, DG-SWEM uses a thin water layer approach detailed in \cite{Bunya2009-jn}. The idea is to apply a ``positive depth" operator $\PD$ to each element at each Runge-Kutta step which prevents that element from having negative depth.
This operator always conserves the total mass of the input element and redistributes surface elevation and flux so that the water depth is greater than a fixed threshold $H_0$ at each node. We then flag the element as ``wet" or ``dry" based on the resulting depth and its subsequent interactions with other elements will depend on this flag.

Specifically, $\PD$ is split into two operators: an operator $\PZ$ applied to water depth and an operator $\PU$ applied to discharge. The characterization of $\PZ$ for each element is as follows. We denote the current depth at node $i$ by $H_i$ and mean depth of element $k$ as $\bar{H}_k$.
\begin{enumerate}
    \item If $H_i \geq H_0 \; \forall i \in \{1,2,3\}$, then $\PZ H_i = H_i$
    \item If $\bar{H}_k \leq H_0$, then $\PZ{H}_i = \bar{H}_k \quad \forall i \in \{1,2,3\}$.
    \item Otherwise, redistribute water mass such that $\PZ H_i \geq H_0 \; \forall i \in \{1,2,3\}$ and $\sum_i \PZ H_i = \sum_i H_i$.
\end{enumerate}
If the updated depth $\PZ H_i$ is less than $H_0$, then node $i$ is flagged as dry.
The discharge operator $\PU$ then removes the momentum from dry nodes and place it on wet nodes so that total momentum is conserved. The exception is when all nodes are dry in which case momentum conservation is allowed to be violated.
\begin{comment}
is specified by:
\begin{align*}
    \PU{p}_i = \Theta_i(p_i + \Delta p /n_{pos})
\end{align*}
where
\begin{align*}
    \Theta_i = 0 \text{ if node } i \text{ is dry}, 1 \text{ otherwise} \\
    n_{pos} = \text{ number of}
\end{align*}
\end{comment}
\begin{comment}
    In the context of our program, the operator is applied at every Runge-Kutta stage:
\begin{align}
    \ww^{(i)}_h := \PD \ww^{(i)}_h
\end{align}
where $\ww^{(i)}_h$ is the state vector during the $i^th$ stage of the SSPRK scheme.
\end{comment}
\begin{comment}
Applying $\PD$ to the solution at each Runge-Kutta stage imposes a restriction on the time step size $\Delta t$ that guarantees. To ensure stability, we also manipulate the numerical flux in the case where the condition is violated:
\begin{align}
    \hat{\FF} = \hat{\FF}(\ww_{in}, \ww_{in}^{ref})
\end{align}
where $\ww_{in}^{ref}$ reflects the normal components of momentum of $\ww_{in}$.
\end{comment}

The thin layer approach is also used in ADCIRC but with a different checking criteria which is nodal-based \cite{Luettich_undated-xy}. A summary of this wet/dry checking sequence at each timestep is as follows:
\begin{enumerate}
    \item Any node with $H < H_0$ is flagged as dry.
    \item For each element that contains one dry node, flag that node as wet if there is sufficient incoming flux from wet nodes connected to it. The threshold for sufficient flux is specified by the user.
    \item Any element that has a positive inflow due to boundary conditions (e.g. normal flow boundary) has all its nodes flagged to wet.
    \item Any node connected to only dry elements is flagged as dry.
\end{enumerate}

Due to these different criteria, we often observe in practice slight differences in wet/dry status between ADCIRC and DG-SWEM at interfaces between wet and dry regions.

\subsection{Incorporating Rainfall Data} \label{sec:rain_models}

To ascertain the rainfall intensity needed in~\eqref{eq:rain}, we essentially have two options: forecast rainfall data from atmospheric models and observed and spatially interpolated rainfall data available post rain or storm event. As an overarching goal of the present work is to develop numerical compound flood models that are capable of forecasting, post event observed data is less relevant. However, for validation and hindcasting purposes, high quality post-event rainfall data is critical.

\begin{figure}[t]
    \centering \includegraphics[width=1.0\textwidth]{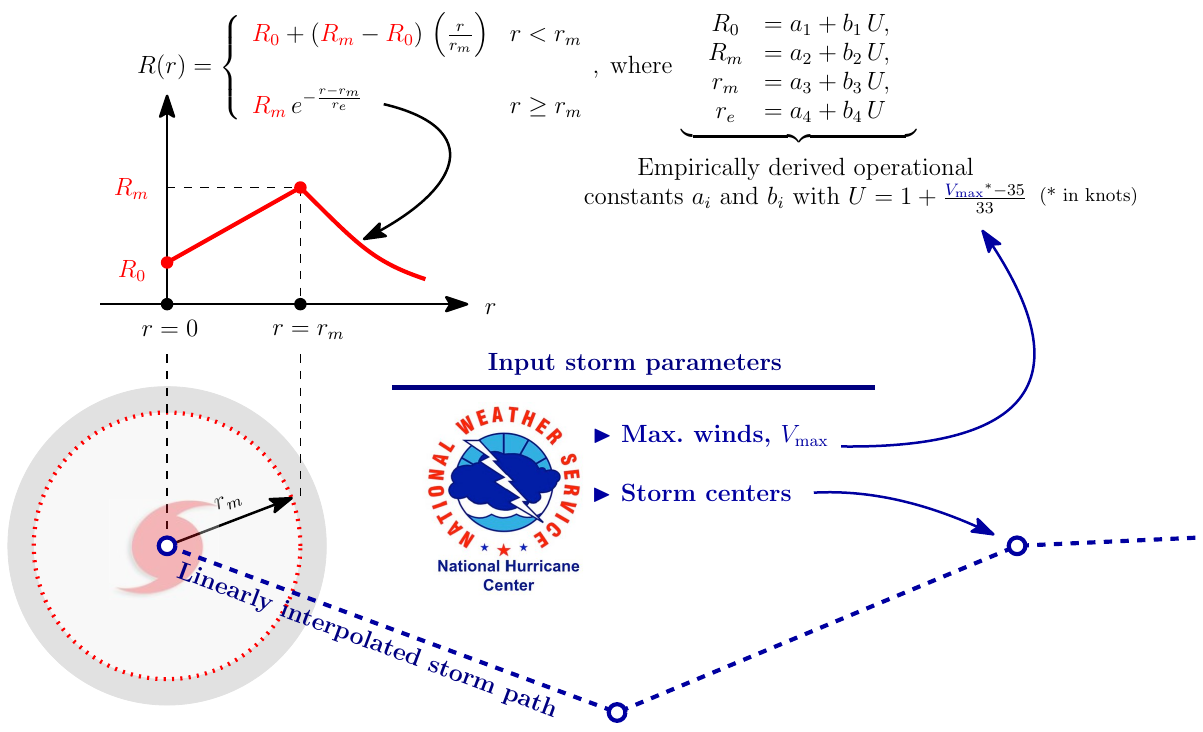}
    \caption{Illustration summarizing the R-CLIPER parametric rainfall model, which computes rainfall rate $R(r)$ at a radius $r$ from a given storm center using the expressions shown. The primary inputs to the model are the reported maximum wind velocities, $V_{\max}$, and the storm centers, which are generally provided at 6-hour intervals by the National Hurricane Center (NHC).}
    \label{fig:RCLIPER}
\end{figure}

During tropical cyclones, rainfall patterns in the vicinity of the storm exhibit defined structures that can be exploited by so-called parametric rainfall models. The basic idea behind this approach is to construct rainfall fields using simple analytic expressions based on a small number of (predicted or observed) storm parameters that are made publicly available by the NHC pre and post storm events.
This type of approach has been implemented in DG-SWEM based on the so-called R-CLIPER (Rainfall CLImatology and PERsistence) model, which computes rainfall rates $R(r)$ at any point in the grid, where $r$ is the distance from that point to the storm center; this is summarized in Figure \ref{fig:RCLIPER}. Here we use the operational $a$ and $b$ parameters given in the bottom four rows of \cite[Table~2]{Tuleya:2007}. Note that the primary inputs of the model are the reported storm centers and maximum wind velocities, which are generally provided at 6-hour intervals by the NHC.
Within DG-SWEM, linear interpolation is used to obtain the storm parameters at the model time step, and the rainfall rate at each finite element mesh node is computed based on its radial distance from the storm center.

Alternatively, the program accepts observed rainfall data in hindcasting scenarios in the GRIB2 format \cite{grib2} which is read periodically based on the time record increment of the data. As this type of data is given with spatially varying distributions over the nation, this data is readily interpolated onto finite element meshes.

\section{Numerical Experiments and Evaluation} \label{sec:experiments}

In this section we present a series of numerical experiments designed to verify and validate the developed numerical model. First, a test case in a simple rectangular geometry which is designed to test the conservation properties of the method. Second, a test case from literature for rainfall onto an inclined plane to compare to laboratory experimental data.
Third, we perform a large-scale test case corresponding to Hurricane Harvey (2017).
Last, an example in Neches River, Texas will be provided that illustrates an alternative compound flood approach commonly used in ADCIRC, where runoff data is incorporated through river boundary conditions.

In all cases, we employ a similar set-up of the solver using Manning's friction law, linear discontinuous approximation functions $(p = 1)$, Roe flux, and a two-stage, second order SSPRK scheme.
The Roe flux is chosen as the default numerical flux because it has been shown to be able to resolve discontinuities in the solution \cite{Cockburn1989-dc}. From our experience, a combination of $p=1$ and second-order time discretization  is sufficiently accurate, and the lowest RK stage for which the simulation is stable is chosen.
The only exception is the Hurricane Harvey case where the mesh is highly resolved. In this case we use the LLF flux as we have found it to be more stable than the Roe flux for this test case at the same time step --- the tradeoff in principle is that the LLF flux is more dissipative at discontinuities \cite{Qiu2006-vq}. To further ensure stability, we also increase the RK stage to 5 which is shown to improve the CFL restriction by the largest amount \cite{Kubatko2014-nc}.

\subsection{Filling a ``Bathtub"}

%\eirik{@Namo, lets show a test case with constant rainfall onto a rectangle domain with a submerged "hump" in the middle. All initial and boundary conditions are zero and lets have rain on top for a day or so to show that the rain does not induce flow when it falls on a flat water surface. This will also show that we satisfy a modified lake-at-rest, see e.g., page 62 in \url{https://repositories.lib.utexas.edu/bitstream/handle/2152/47014/ARABSHAHI-DISSERTATION-2016.pdf?sequence=1}}

\begin{figure}[h]
    \centering
    \includegraphics[width=0.85\textwidth]{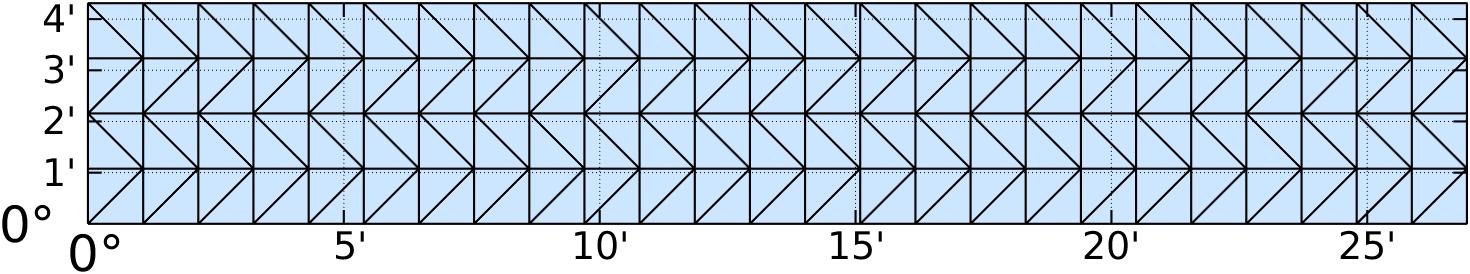}
    \caption{Bathub mesh. Coordinates are in degrees and decimal minutes format.}
    \label{fig:tubmesh}
\end{figure}

\begin{figure}[h]
    \centering
    \includegraphics[width=0.95\linewidth]{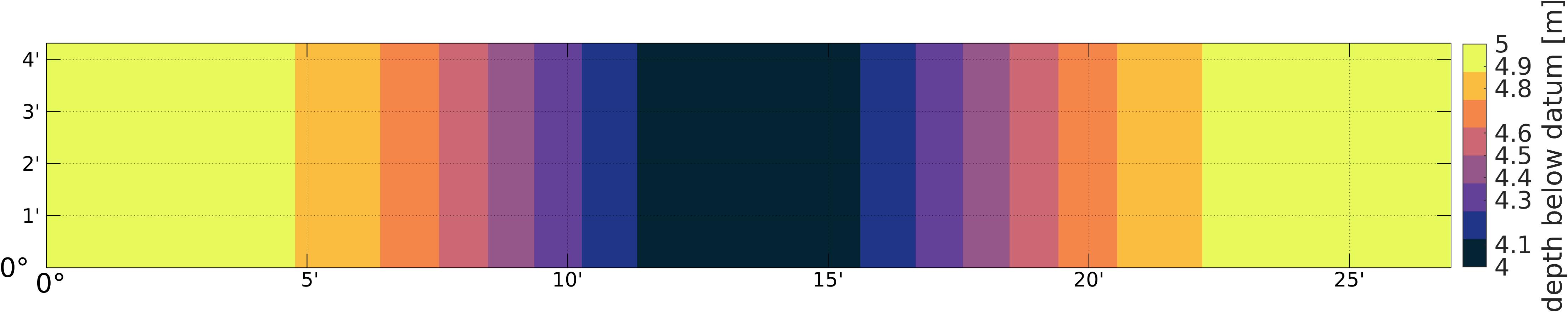}
    \caption{Bathymetry of the bathtub mesh.}
    \label{fig:tub}
\end{figure}

As an initial numerical experiment we verify the well-balanced nature of our DG solver in the presence of nonzero source terms. We accomplish this by considering a modification of the classical lake-at-rest test case~\cite{leveque1998balancing}. Hence, we consider a rectangular domain $(0,50 000m)\times(0,8 000m)$ and consisting of 200 triangular elements (covered by 25 in the horizontal direction, 4 in the vertical), see Figure \ref{fig:tubmesh}. As shown in Figure
\ref{fig:tub}, the bathymetry is set to be
\begin{equation} \label{eq:bathy1}
    h_b(x,y)=    5  -  e^{-100(x-25000m)^2},
\end{equation}
creating a ``bump" at the center.

We assume that the bottom has a constant quadratic friction coefficient of 0.001, set the time step size to $5$ seconds, and apply zero-flow boundary conditions on all domain edges.
We incorporate a constant rainfall rate of $7.0556 \times 10^{-6}$ m/s (1 inch/hour) for 24 hours, followed by 48 hours of no forcing.
At the conclusion of the simulation, the water surface elevation is reported to be $0.6096$ m, an exact match when computing the volume added by the constant rainfall. Lastly, the maximum velocity throughout the simulation is $0.35 \times 10^{-13}$ m/s. Hence, we conclude that the inclusion of rainfall into the DG solver does not disturb the well-balanced property.

\subsection{Rain on Inclined Plane}

%\eirik{@Ethan can you add the Iwagaki stuff here?}
With this test case, we validate the ability of DG-SWEM to accurately simulate rainfall runoff on an (initially dry) inclined plane subjected to rainfall rates of varying duration. Specifically, computed water depths and outflow discharges are validated against a set of well-known experimental results from Iwagaki \cite{Iwagaki:1955}, which have been used in a number of validation studies; see, for example, \cite{Feng:1997,Santillana:2010,Zhang:1989}.

\begin{figure}[h]
    \centering    \includegraphics[width=1.0\textwidth]{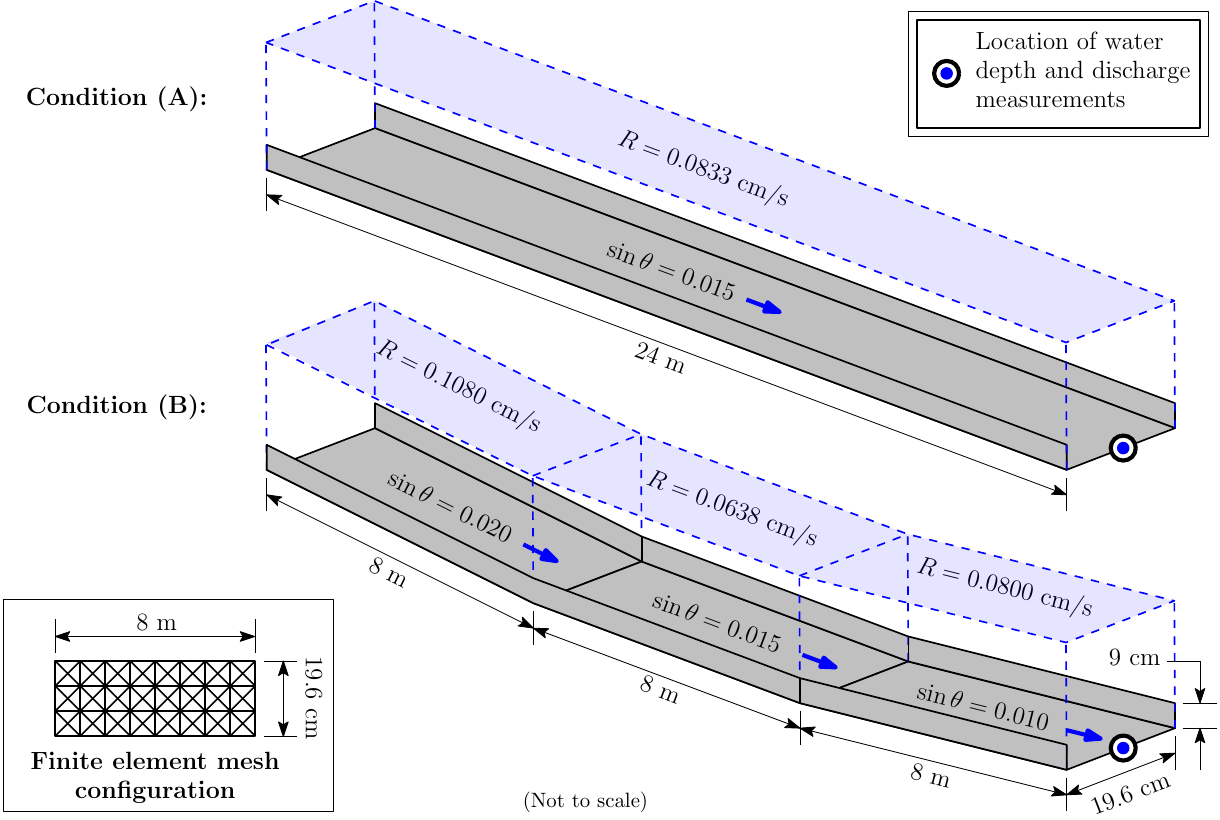}
    \caption{Iwagaki's experimental setup for conditions (A) and (B) with rainfall rates, $R$, as shown for time durations of $T=10$, $20$, and $30$ seconds. Bottom inset shows the finite element mesh configuration used for simulation.}
    \label{fig:IwagakiSchematic}
\end{figure}

Figure \ref{fig:IwagakiSchematic} shows the two experimental ``conditions'' investigated by Iwagaki, both of which used a 24 m-long aluminum flume having a rectangular cross section of 19.6 cm width and 9 cm depth. In condition (A), the flume was set at a uniform slope of $\sin \theta = 0.015$ and subjected to a uniform rainfall rate of $R=0.0833$ cm/s. In condition (B), the flume was set (from top to bottom) at slopes of $\sin \theta = 0.020$, $0.015$, and $0.010$ and subjected to rainfall rates of $R=0.1080$ cm/s, $0.0638$ cm/s, and
$0.0800$ cm/s, respectively, with each segment being 8 meters long. For both conditions, rainfall rates of time duration $T=10$, $20$, and $30$ seconds were investigated and measurements of the water depth and discharge were reported at the end of the flume.

\begin{figure}[H]
    \centering
    \includegraphics[width=0.49\textwidth]{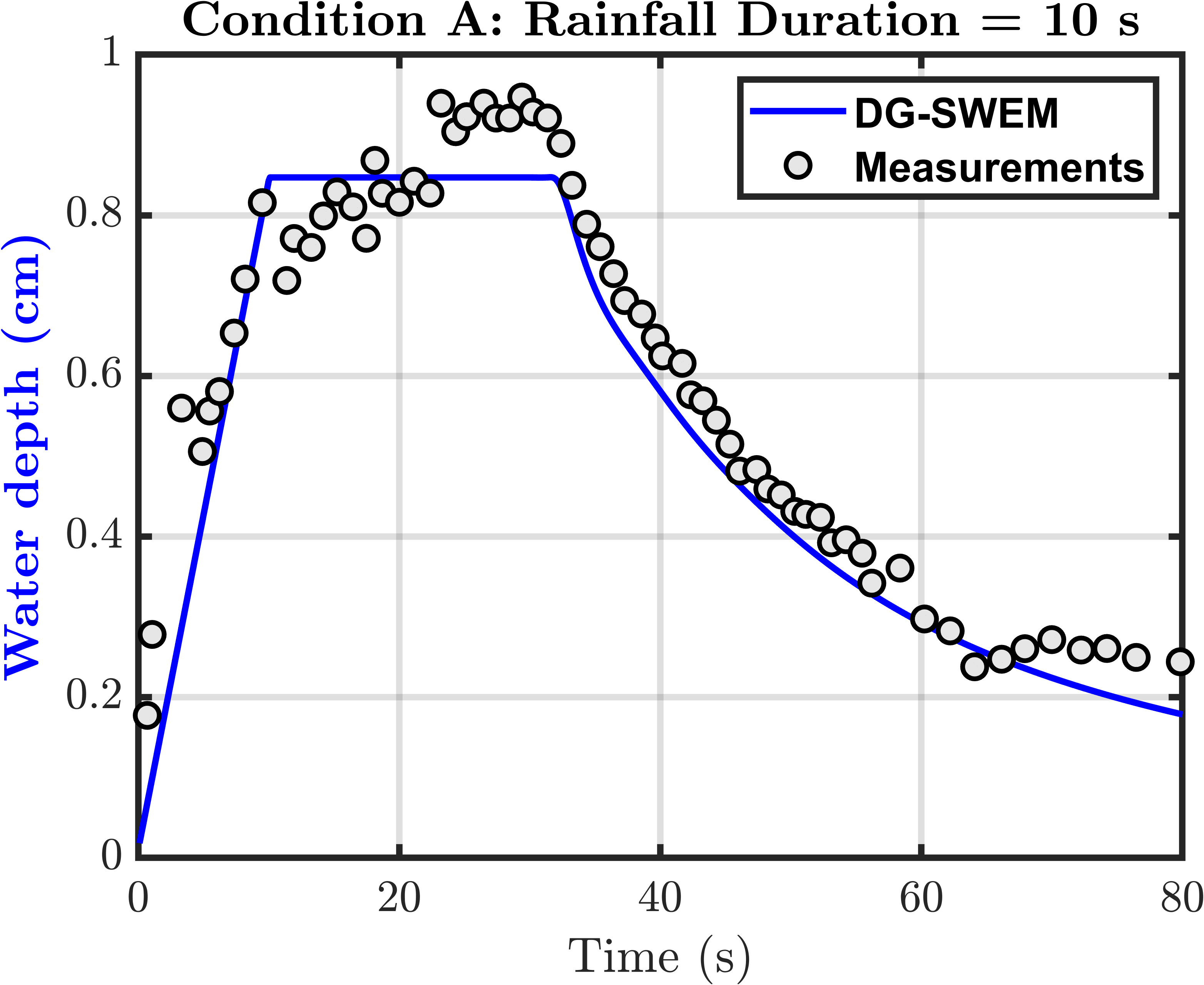}
    \includegraphics[width=0.49\textwidth]{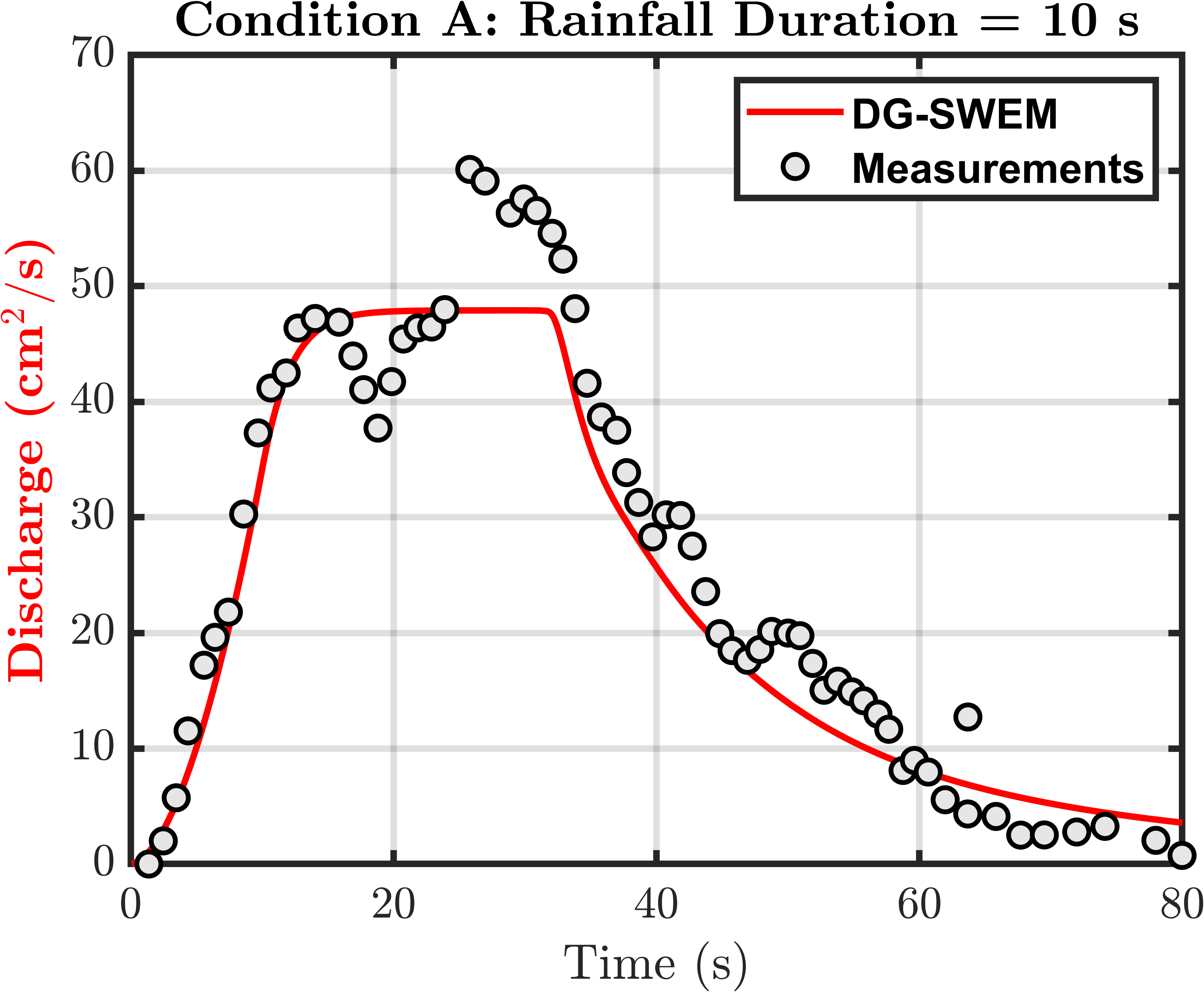}\\
    \vspace{0.10in}
    \includegraphics[width=0.49\textwidth]{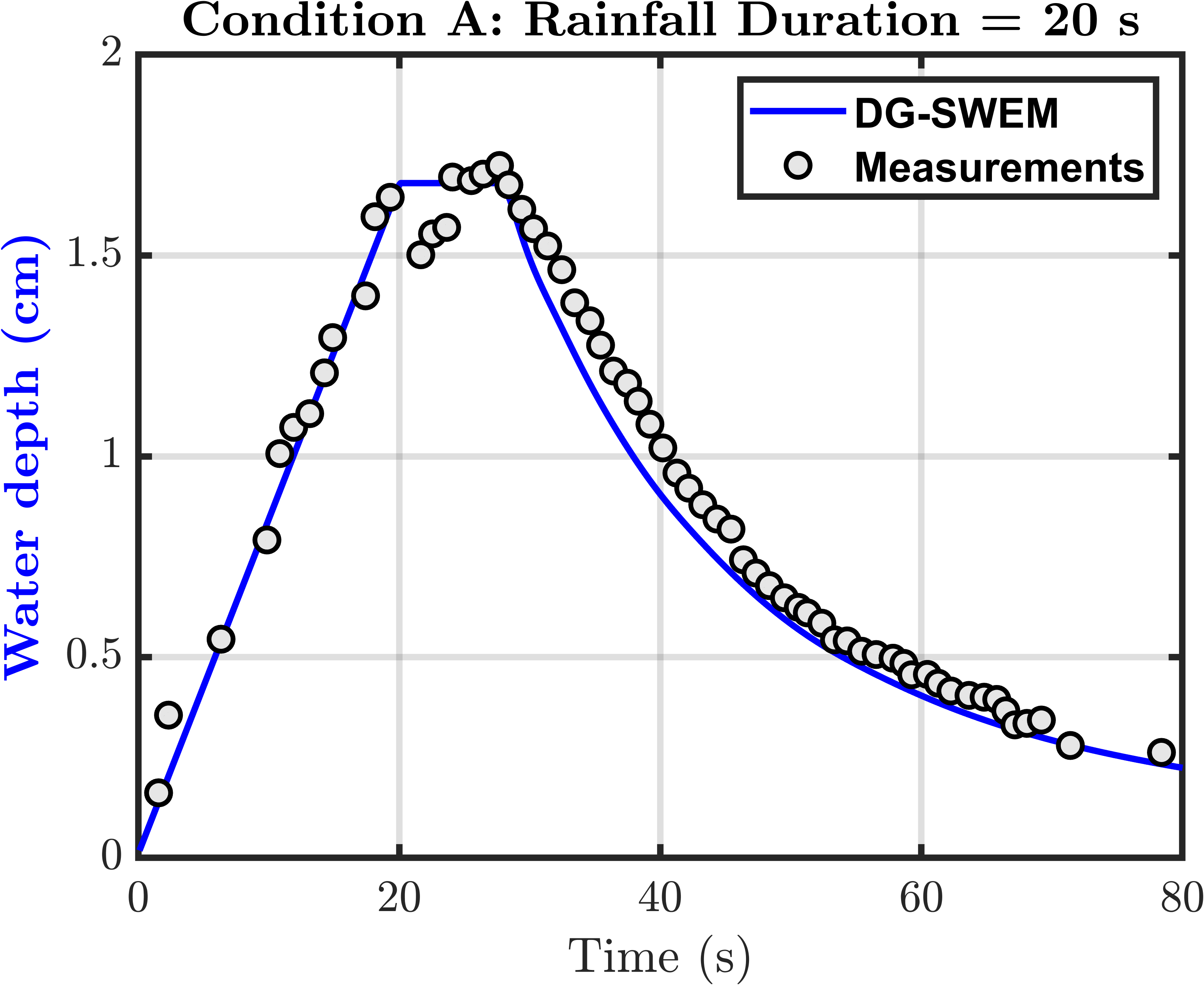}
    \includegraphics[width=0.49\textwidth]{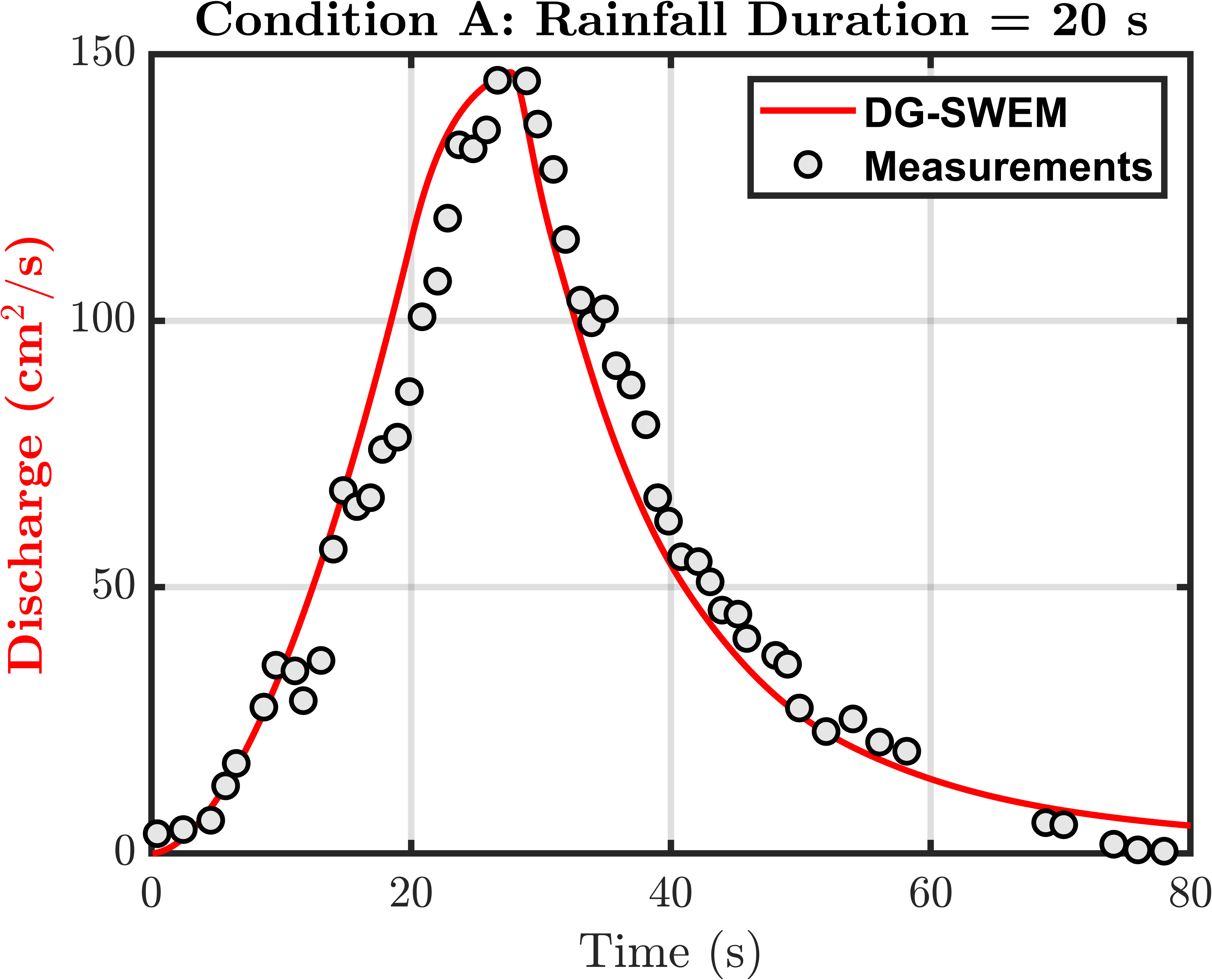}\\
    \vspace{0.10in}
    \includegraphics[width=0.49\textwidth]{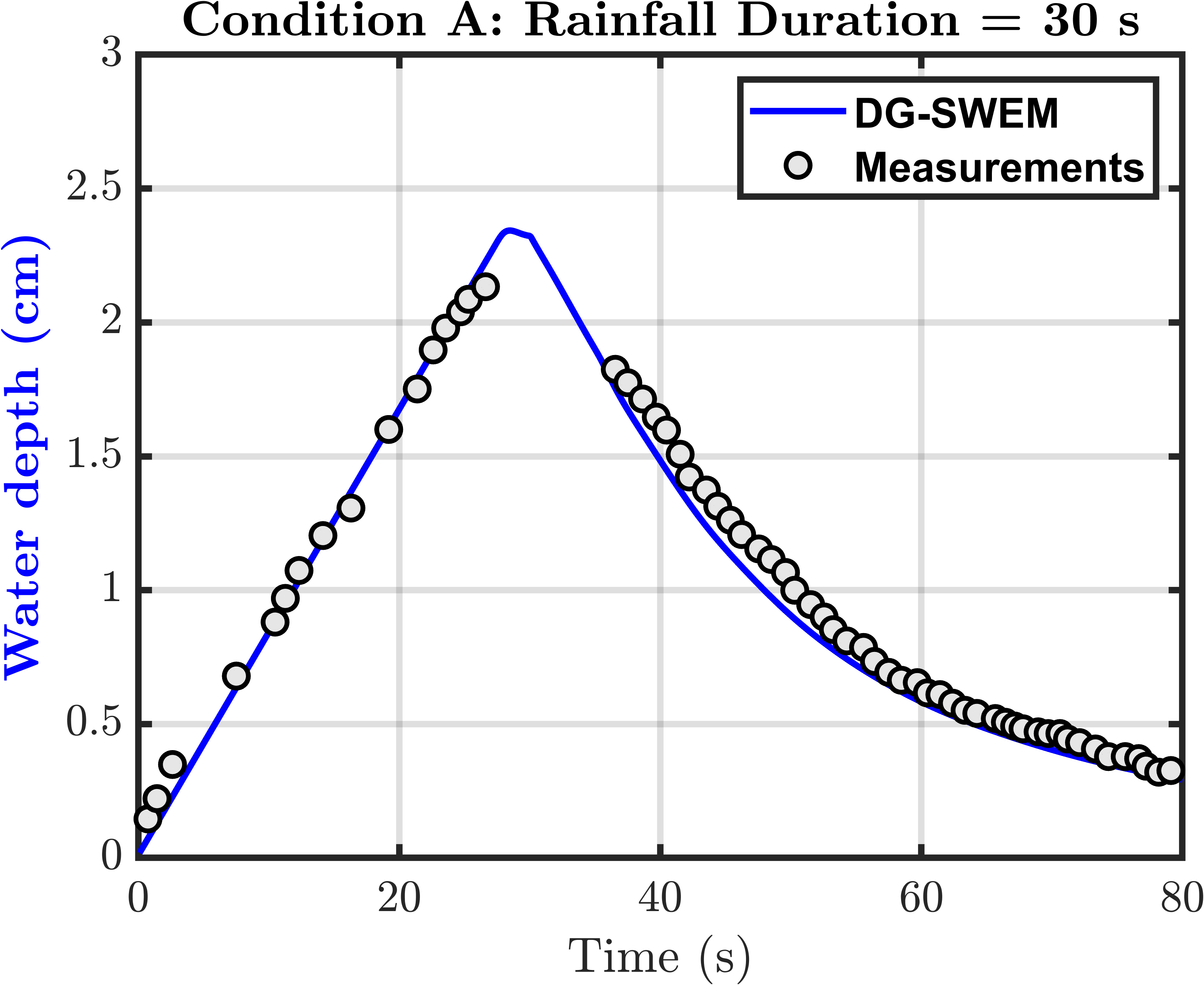}
    \includegraphics[width=0.49\textwidth]{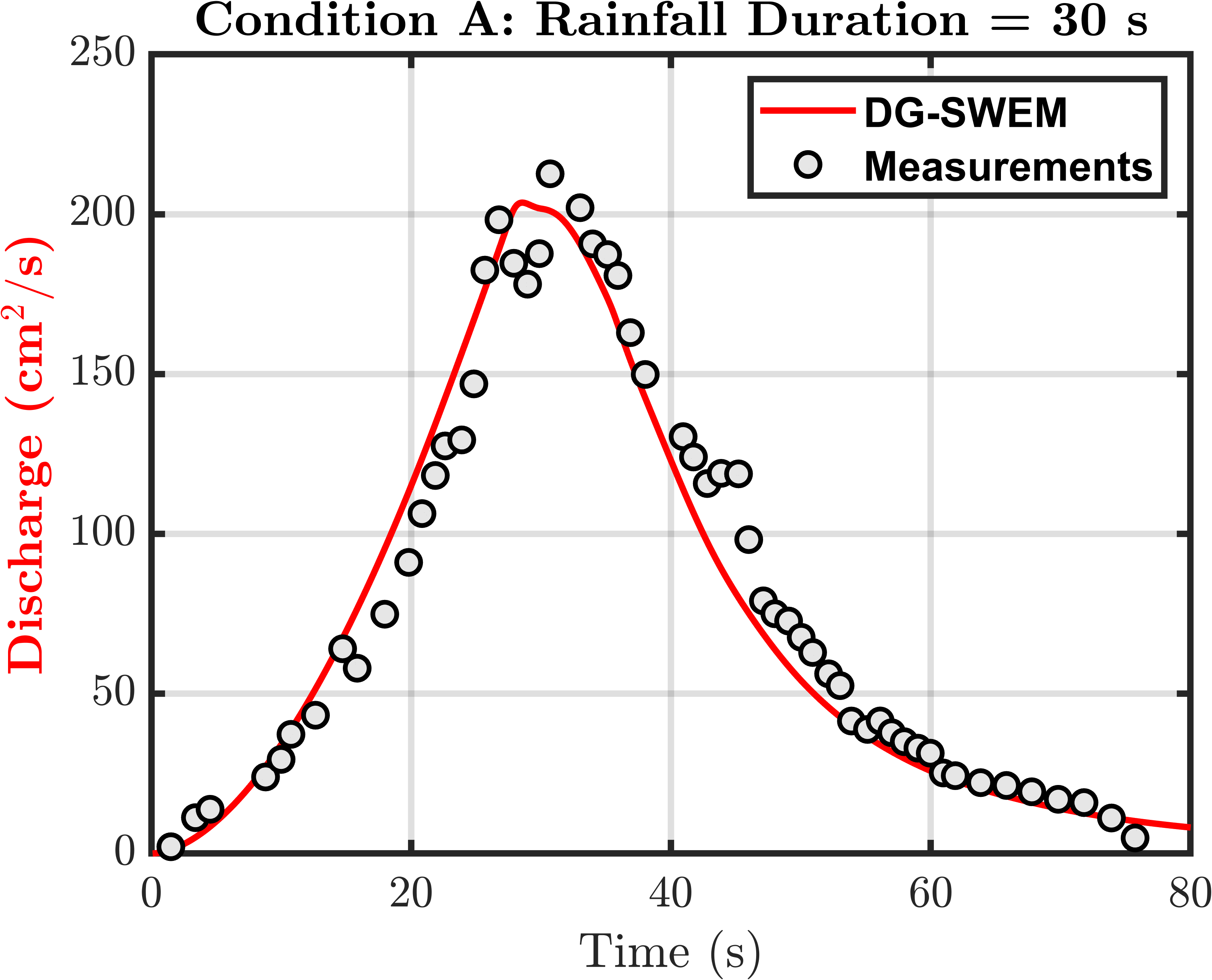}
    \caption{Plots for Iwagaki's ``Condition (A)'' comparing DG-SWEM results for water depth (\textit{left, blue}) and discharge (\textit{right, red}) to Iwagaki's measurements (\textit{gray circles}) at the end of the 24-m flume. Rainfall durations are 10, 20, and 30 seconds (\textit{top}, \textit{middle}, and \textit{bottom}, respectively).}
    \label{fig:IwagakiReultsA}
\end{figure}

\begin{figure}[H]
    \centering
    \includegraphics[width=0.49\textwidth]{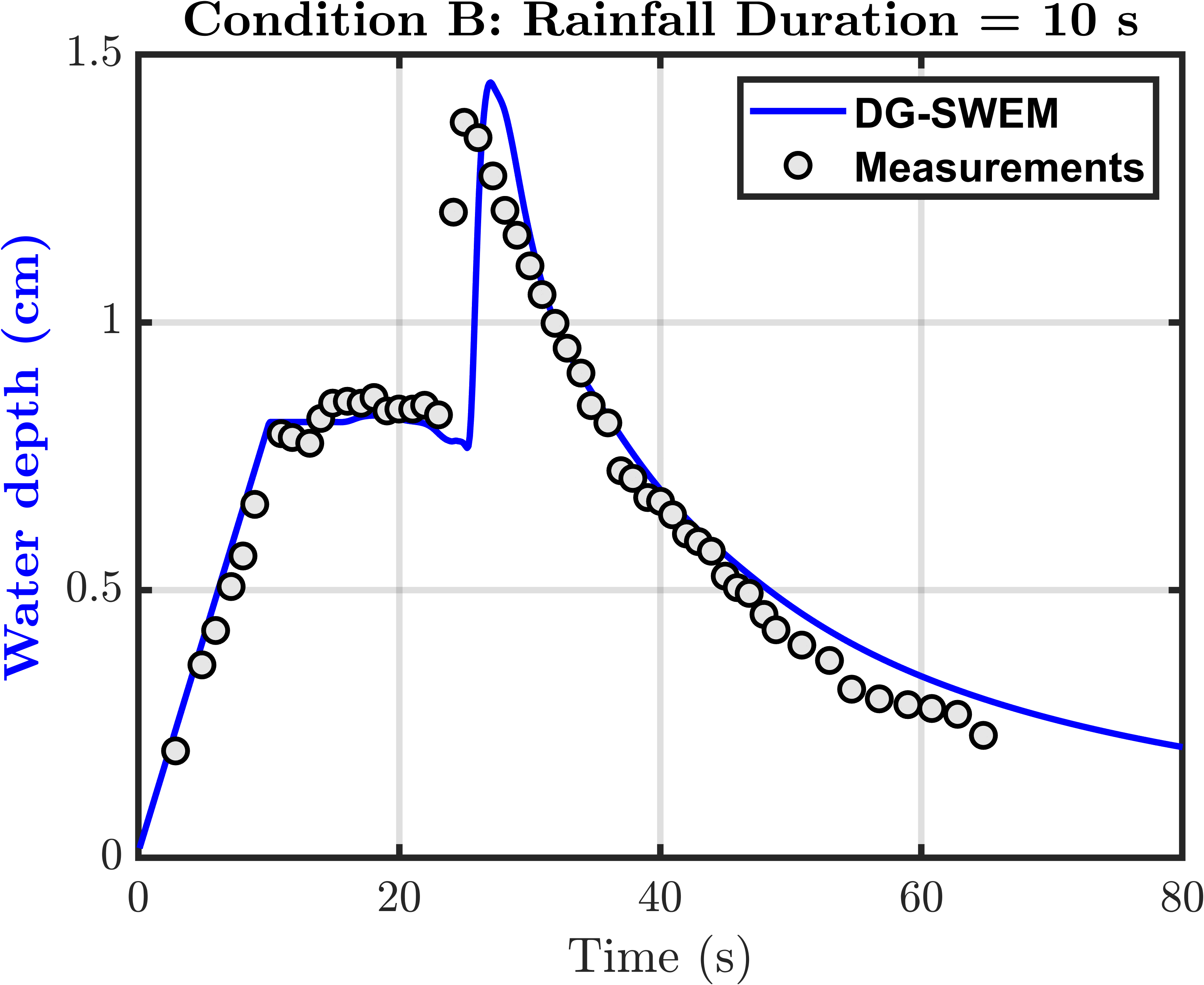}
    \includegraphics[width=0.49\textwidth]{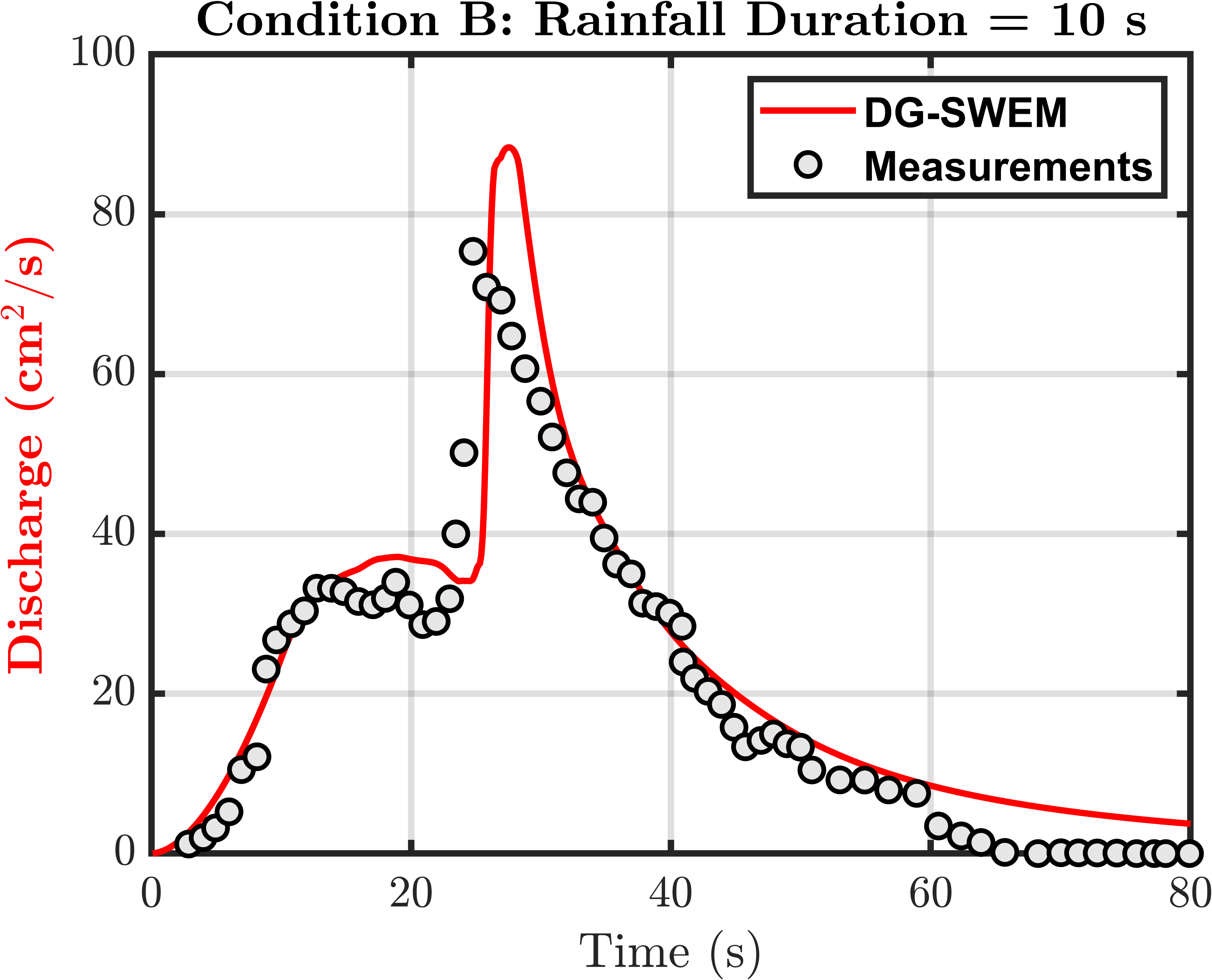}\\
    \vspace{0.10in}
    \includegraphics[width=0.49\textwidth]{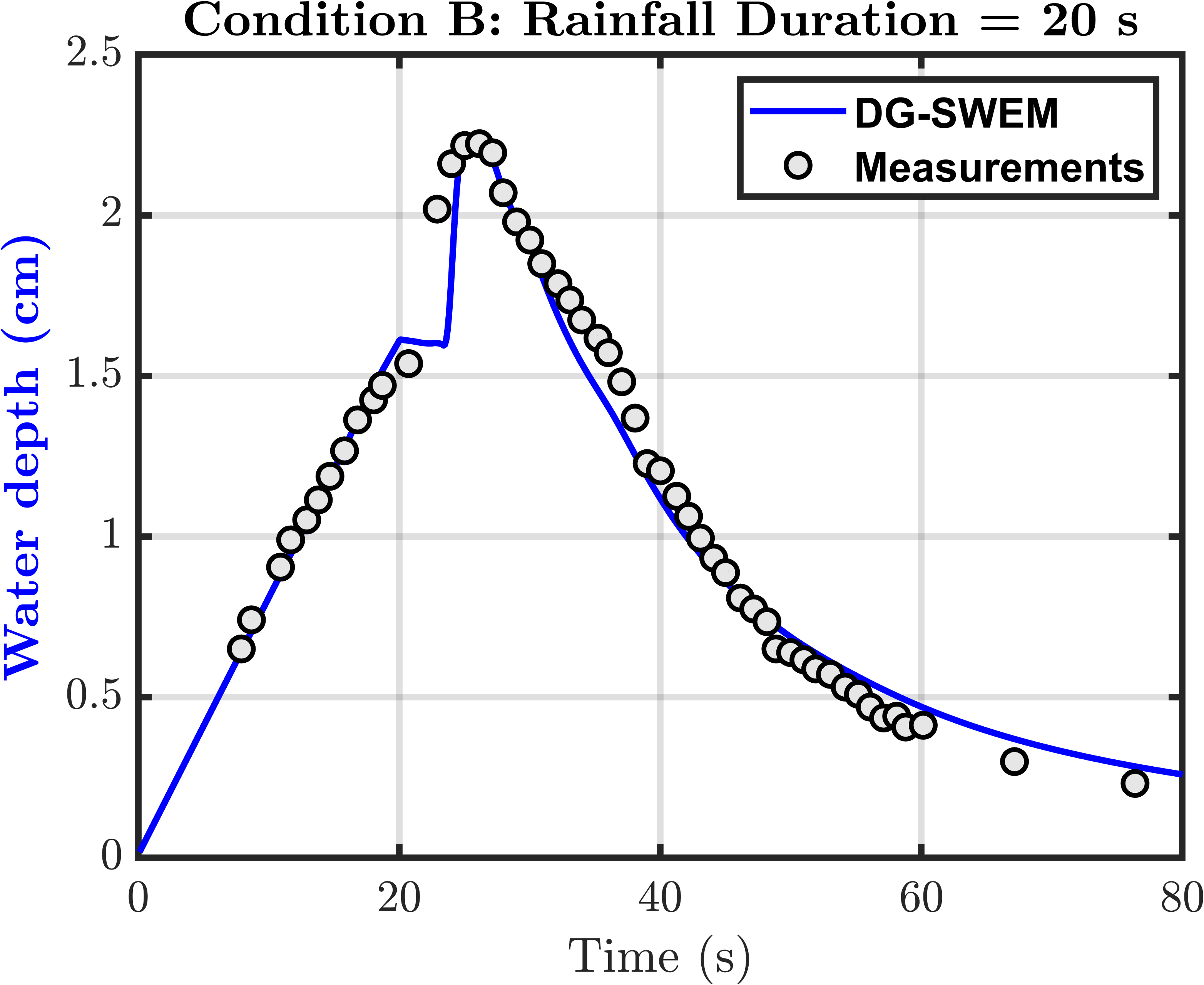}
    \includegraphics[width=0.49\textwidth]{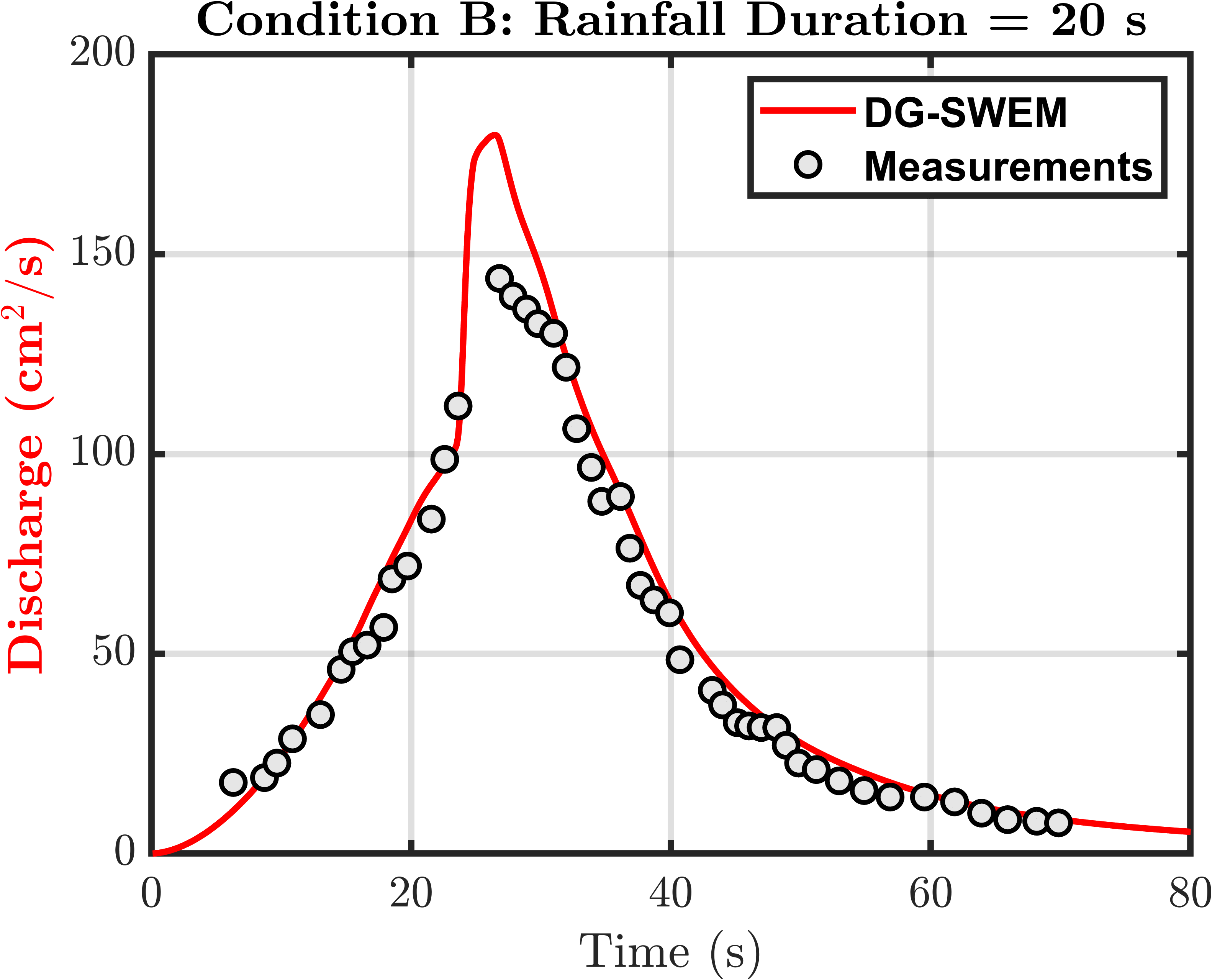}\\
    \vspace{0.10in}
    \includegraphics[width=0.49\textwidth]{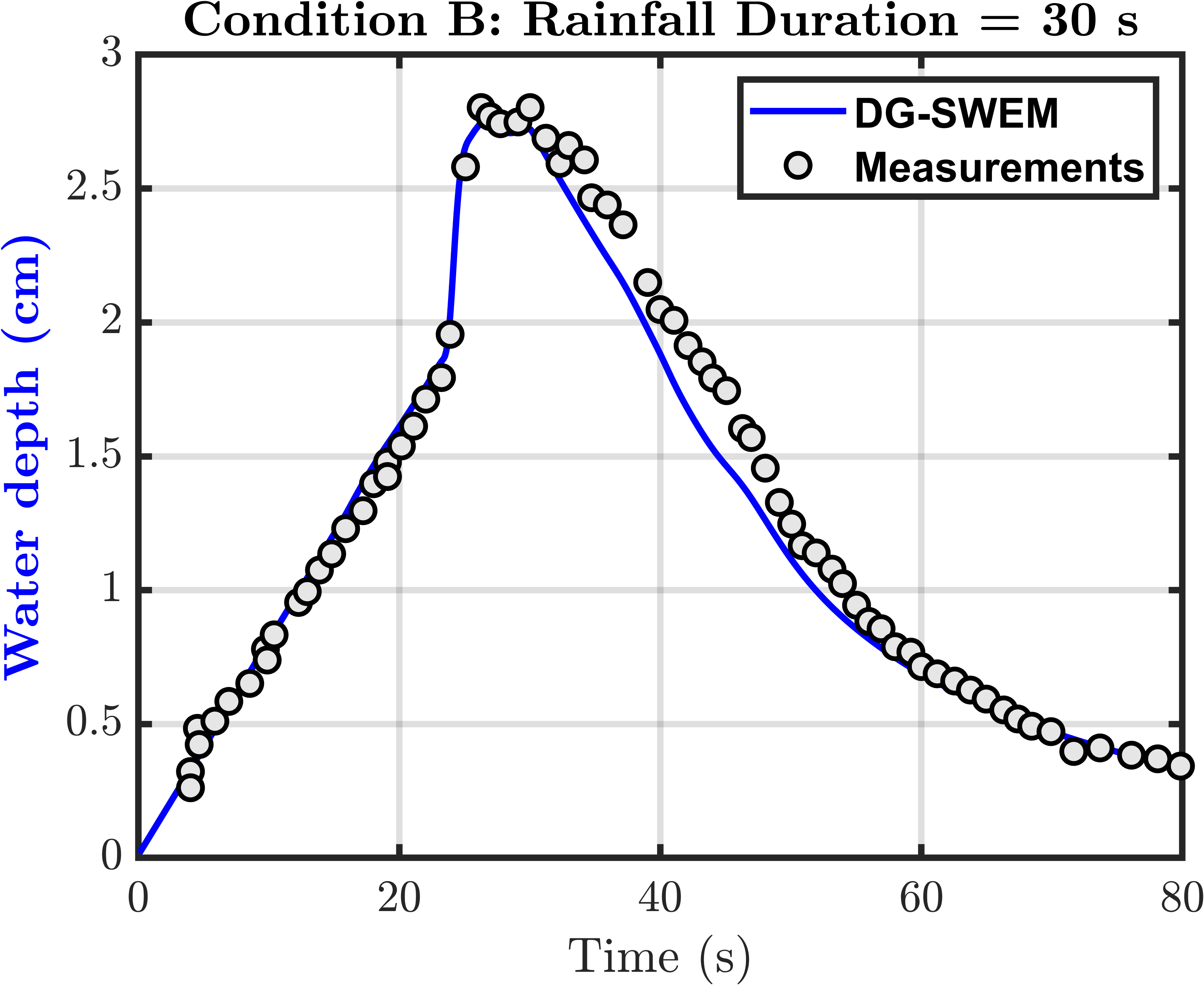}
    \includegraphics[width=0.49\textwidth]{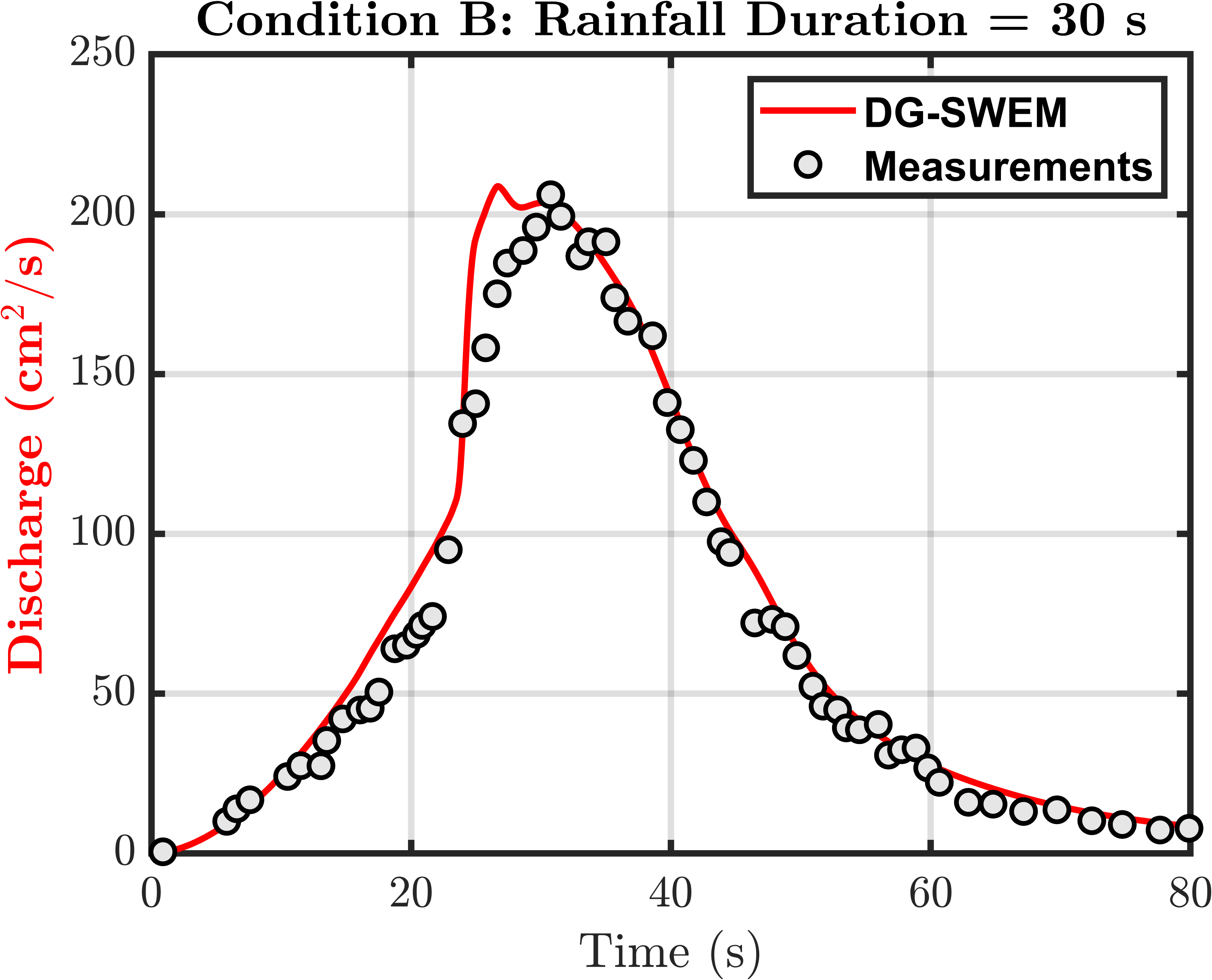}
    \caption{Plots for Iwagaki's ``Condition (B)'' comparing DG-SWEM results for water depth (\textit{left, blue}) and discharge (\textit{right, red}) to Iwagaki's measurements (\textit{gray circles}) at the end of the 24-m flume. Rainfall durations are 10, 20, and 30 seconds (\textit{top}, \textit{middle}, and \textit{bottom}, respectively).}
    \label{fig:IwagakiReultsB}
\end{figure}

Each 8-m segment of the flume is discretized as shown in the inset of Figure \ref{fig:IwagakiSchematic}. Numerical simulations for both conditions (A) and (B) were performed with DG-SWEM for each of the three rainfall durations, $T=10$, $20$, and $30$ seconds, using piecewise polynomial spaces of degree $p = 1$. As reported by Iwagaki, Manning's coefficient was set to $n=0.009$ (no model calibration was performed), and numerical simulations were run for a total of 86.4 seconds using a time step of $\Delta t = 0.001$ seconds. Figures \ref{fig:IwagakiReultsA} and \ref{fig:IwagakiReultsB} show the computed water depth and outflow discharge at the end of the flume obtained from the DG-SWEM simulation compared to Iwagaki's experimental results for conditions (A) and (B), respectively. Overall, it can be observed that the DG-SWEM results closely match the experimental measurements, with computed output accurately reproducing the rising and recession limbs of the hydrographs. The disagreements we observe are less than $0.2$ cm in the elevation, and $25$ cm$/$s in the discharge.
Additionally, in general, the peak water depth and discharge values, and the time at which these peaks occur, are well captured with the model. For example, the root-mean-square error in peak water depths under condition (B) is only 0.06 cm.
\newpage
\newpage

\subsection{Hurricane Harvey}
\begin{figure}[H]
\centering
\includegraphics[width=0.60\textwidth]{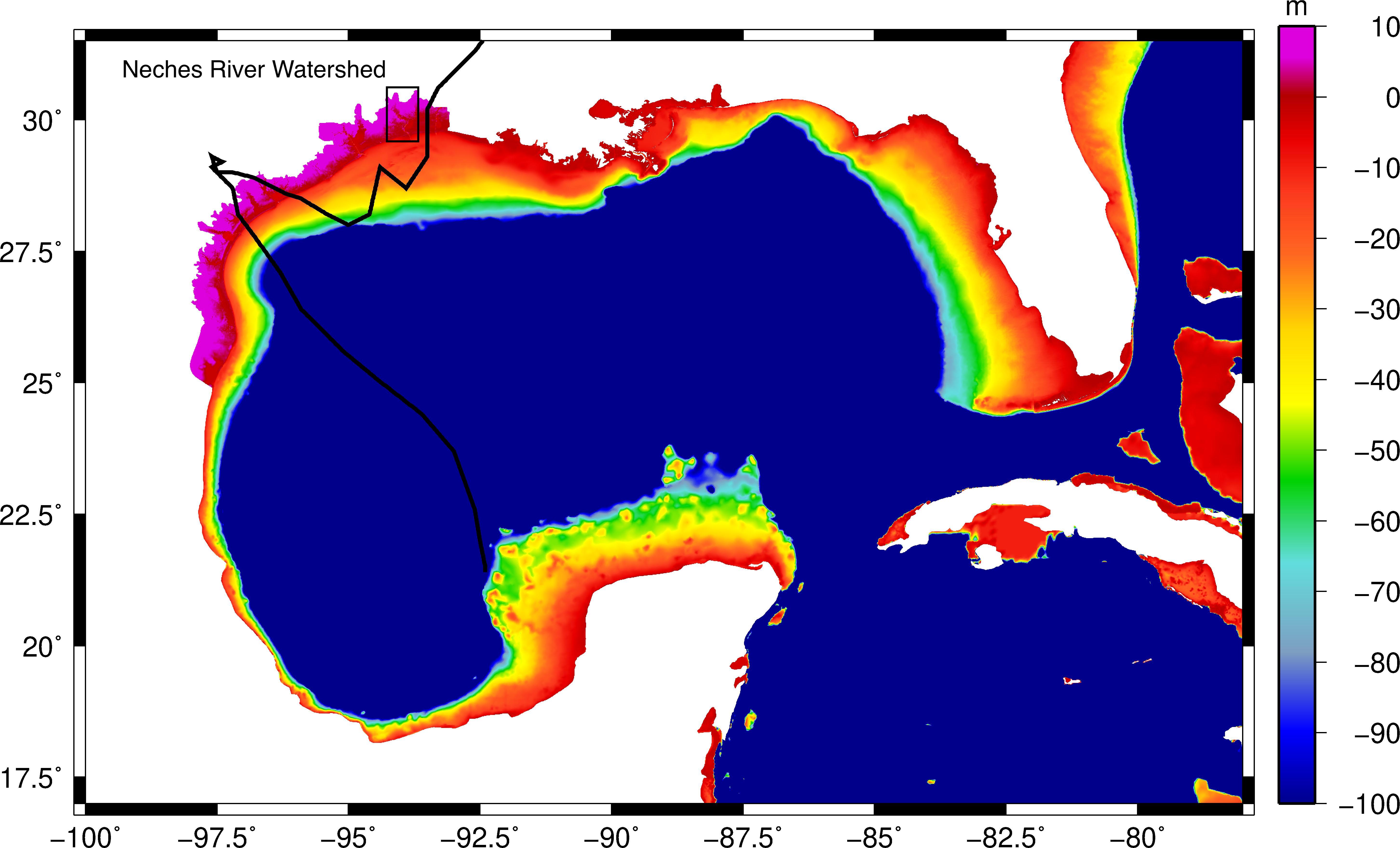}
\caption{\label{fig:harvey_track} Track of Hurricane Harvey in the Western Gulf of Mexico. Note that the color contours represents bathymetry. }
\end{figure}

As a final validation test, we consider a canonical case of compound flooding, Hurricane Harvey (2017). During the storm, moderate surge in Galveston Bay  compounded with more than 50 inches of total rain in the Houston area~\cite{valle2020compound} and induced historic floods in the Houston area. Hurricane Harvey made its initial landfall on August 25$^{th}$ 2017 on San Jos\'e island~\cite{blake2018hurricane}, before moving back into the Gulf of Mexico and making its second landfall near the Texas-Louisiana border two days later, see Figure~\ref{fig:harvey_track} for its track in the Gulf of Mexico. Between these landfalls, the storm system moved back into the Gulf of Mexico and lead to extreme rainfall in Houston and surrounding areas.

\begin{figure}[H]
    \centering
    \includegraphics[width=0.49\textwidth]{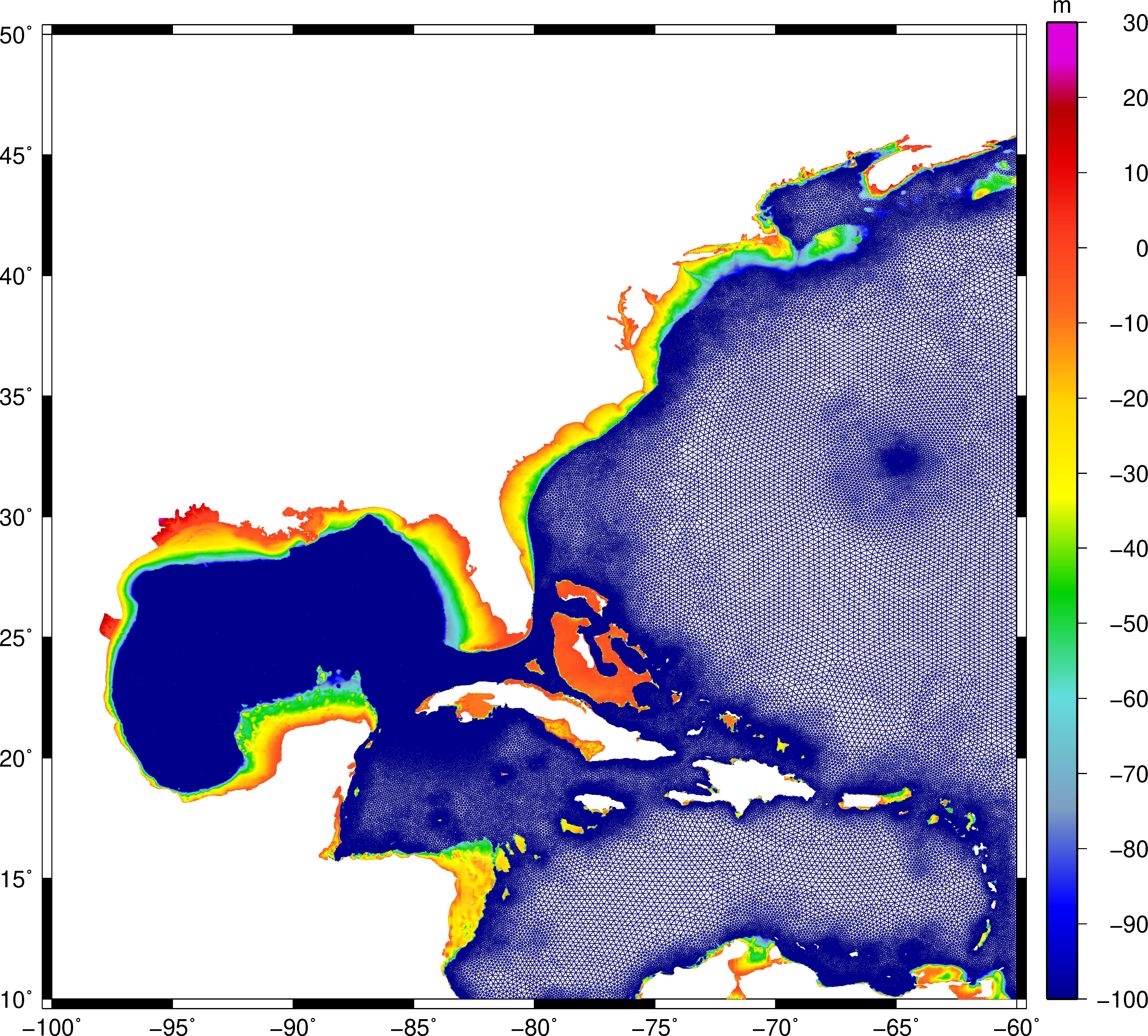}
    \includegraphics[width=0.49\textwidth]{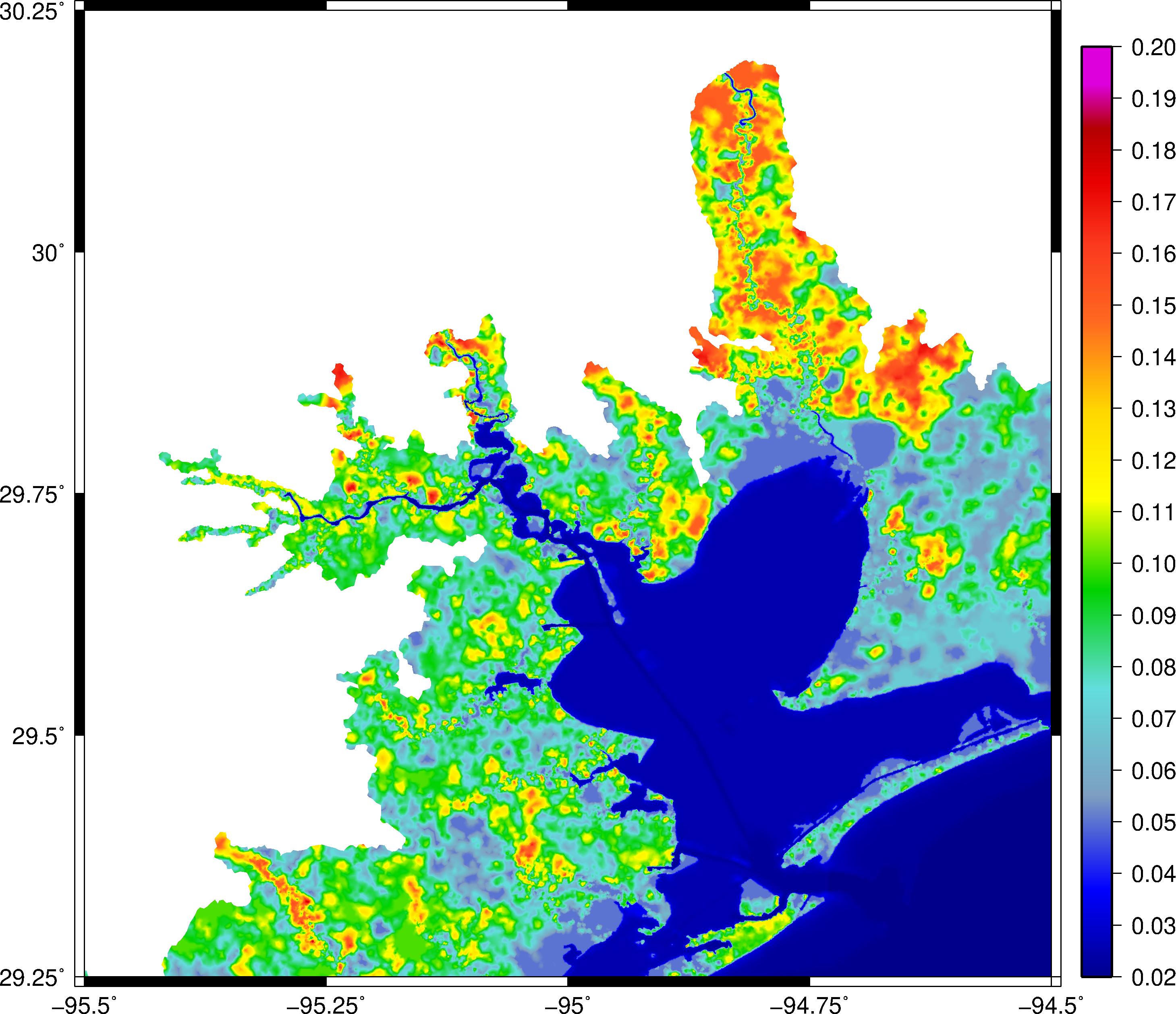}

    \caption{(Left) Bathymetry of the mesh. (Right) Zoomed-in portion of the mesh showing Manning's n values in the Galveston Bay area.}
    \label{fig:bathy}
\end{figure}

To perform simulations of the floods during Hurricane Harvey using the developed DG solver, we rely on established and validated inputs for hurricane storm surge modeling with ADCIRC as described in~\cite{goff2019outflow}. The inputs include an unstructured mesh consisting of 3,352,598 nodes distributed among 6,675,517 triangular finite elements, shown in Figure~\ref{fig:bathy}, as well as spatially variable descriptions of Manning's $n$ \cite[p.~20-22]{Contreras2023-qi}.
The resolution for this mesh is 20 meters for the smallest element. In addition to bathymetry and bottom friction, the inputs also include parametrizations and descriptions of tides, levees, and wind-inhibiting vegetation along the Texas coast.
%
%
%
%\begin{figure}[H]
 %   \centering
  %  \includegraphics[width=0.5\textwidth]{120m_cut_man.jpg}
   % \caption{Zoomed-in portion of the mesh showing Manning's n values in the Galveston Bay area.}
   % \label{fig:mannings}
%\end{figure}
%
%

To emulate the process and data stream in a forecasting scenario, we obtain Best Track data from the National Hurricane Center HURDAT2 database~\cite{landsea2013atlantic}.
%within a parametric hurricane vortex model to generate wind forcing, the Generalized Asymmetric Holland Model (GAHM). This parametric model is based on the classical Holland Model (HM) \cite{holland1980analytic} and has been specifically developed for ADCIRC.
As in~\cite{goff2019outflow}, we perform a tidal spinup without winds to ensure stability and accuracy of the model once hurricane strength winds are applied. The 30-day spinup is followed by a run  from August 23 to September 2, 2017 during which Harvey made landfall, and hurricane data (wind and pressure) and rainfall are incorporated using the GAHM and the R-CLIPER model, respectively.
To ensure stability of the temporal discretization we use a timestep of 0.5 seconds. This fine resolution in both space and time leads to a runtime of approximately 13 hours with 3,200 processors on the Frontera supercomputer at the Texas Advanced Computing Center.

\begin{figure}[h]
    \centering
    \includegraphics[width=0.99\textwidth]{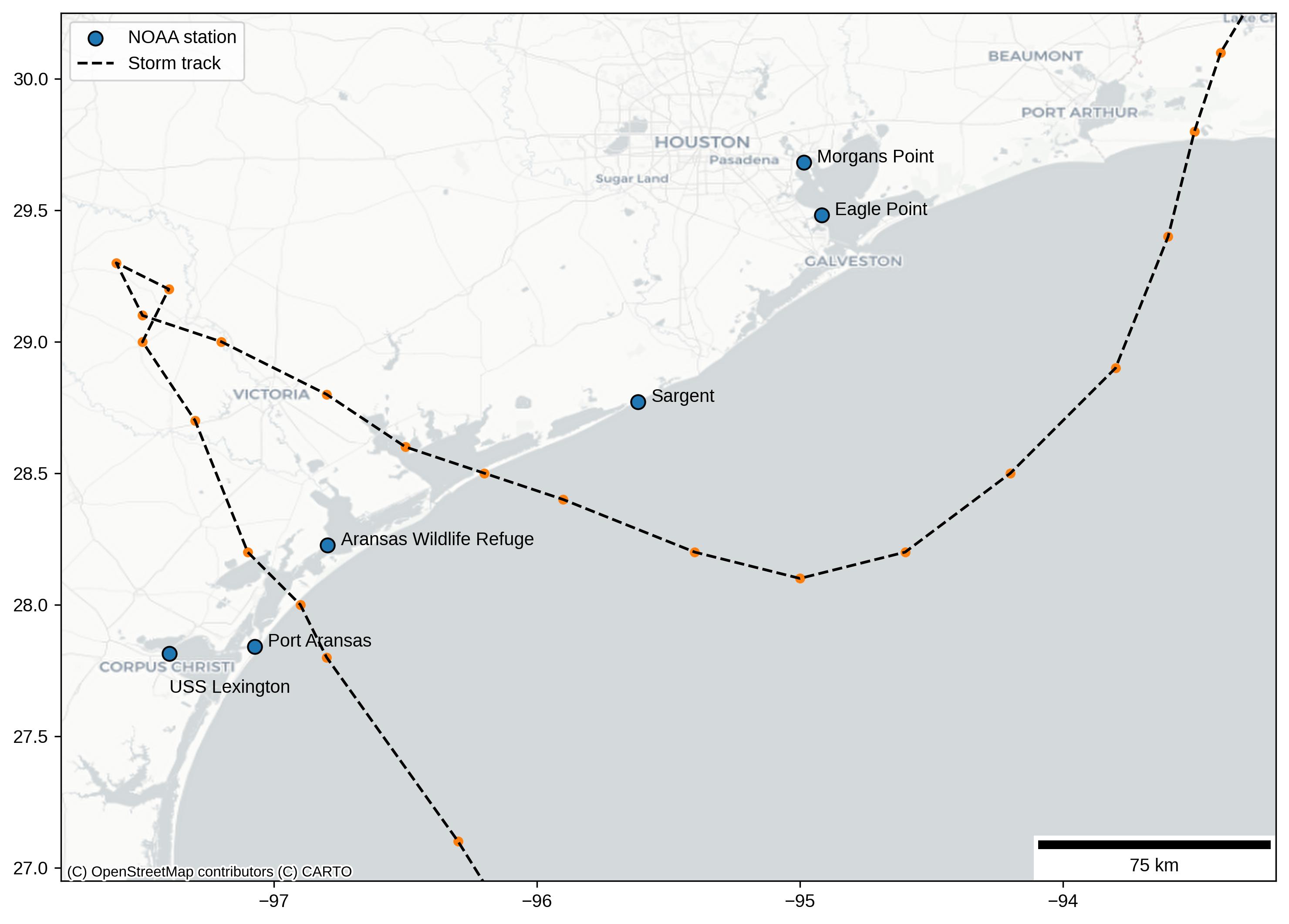}
    \caption{Track of Harvey near the Texas coast. Also shown are the six NOAA water level stations used in this paper.}
    \label{fig:stations}
\end{figure}

To compare the results, we consider three simulation scenarios: 1) a surge-only case with no rainfall, 2) a case with surge and rain from the R-CLIPER model, and 3) a case with surge and observed rain intensity  obtained from Iowa State weather data archive \cite{iowa}. This gridded data is linearly interpolated onto each node and then averaged over each element as in equation \eqref{eq:rain}. To compare our results with observations near the area where Harvey made its initial Texas landfall, we use the NOAA gauge stations shown in Figure \ref{fig:stations}.
These were selected due to their proximity to the location of Hurricane Harvey's first landfall on the Texas coast and where significant compound flooding was observed.

\begin{figure}[htbp]
    \centering
    \includegraphics[width=0.49\textwidth]{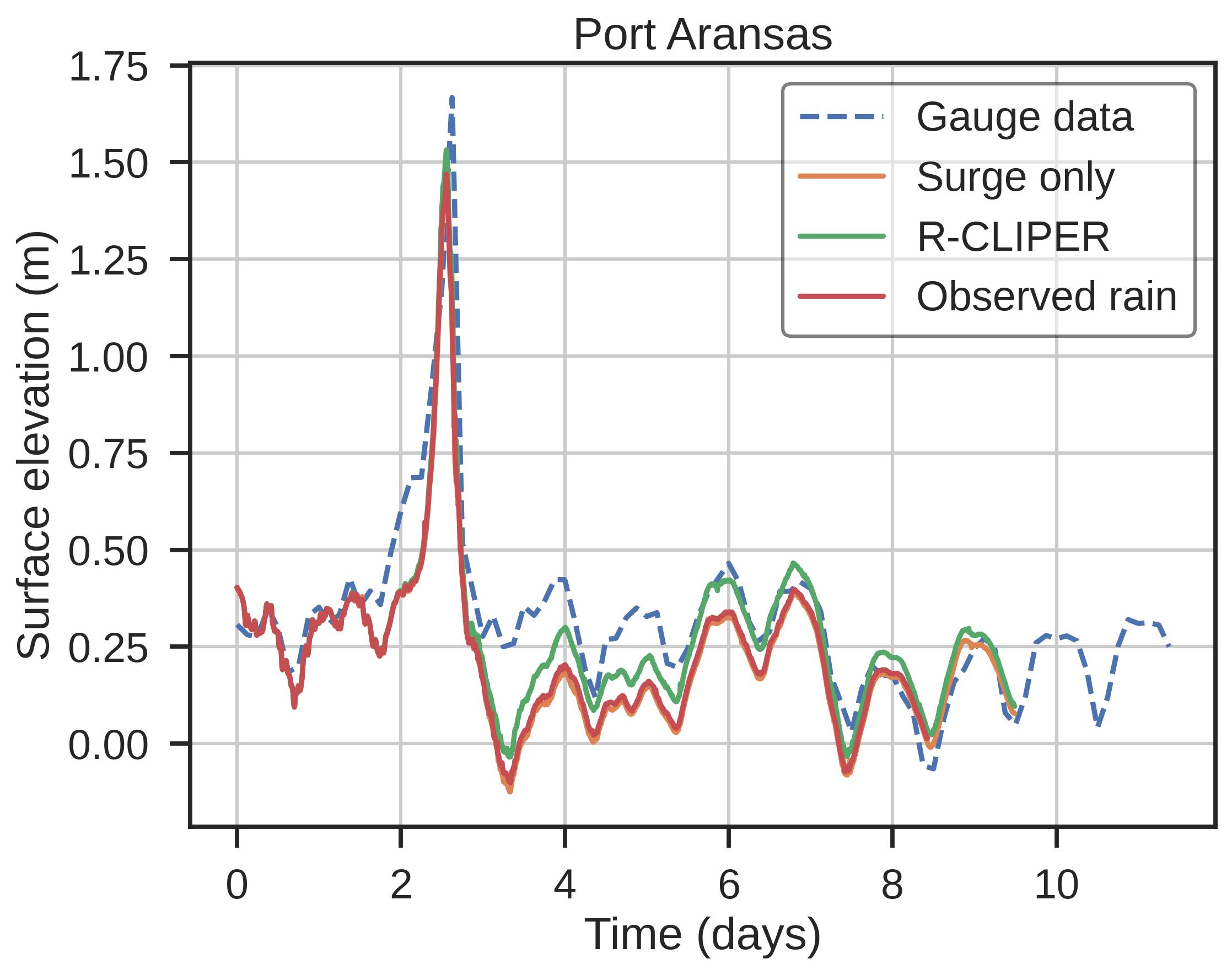}
    \includegraphics[width=0.49\textwidth]{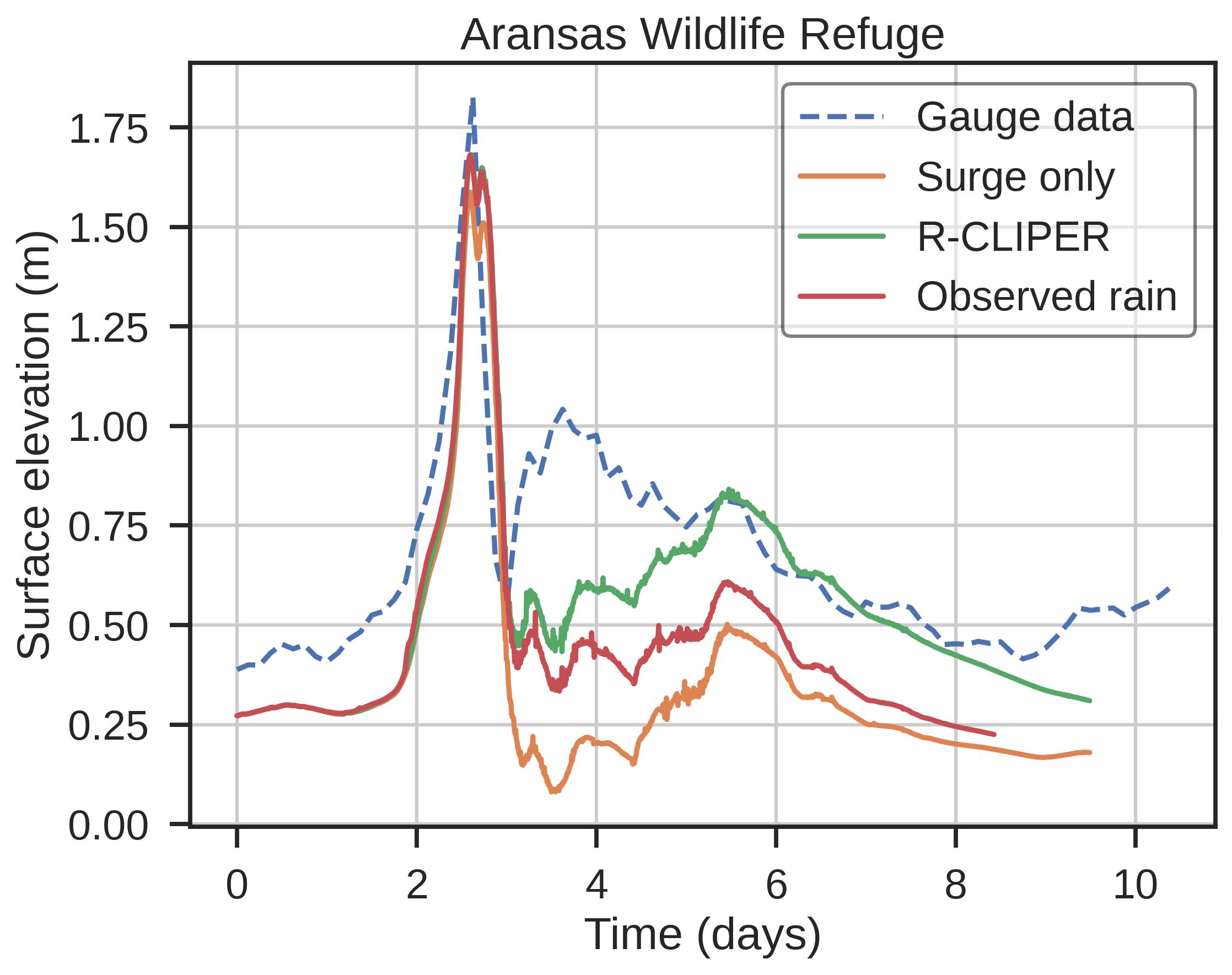}\\
    \includegraphics[width=0.49\textwidth]{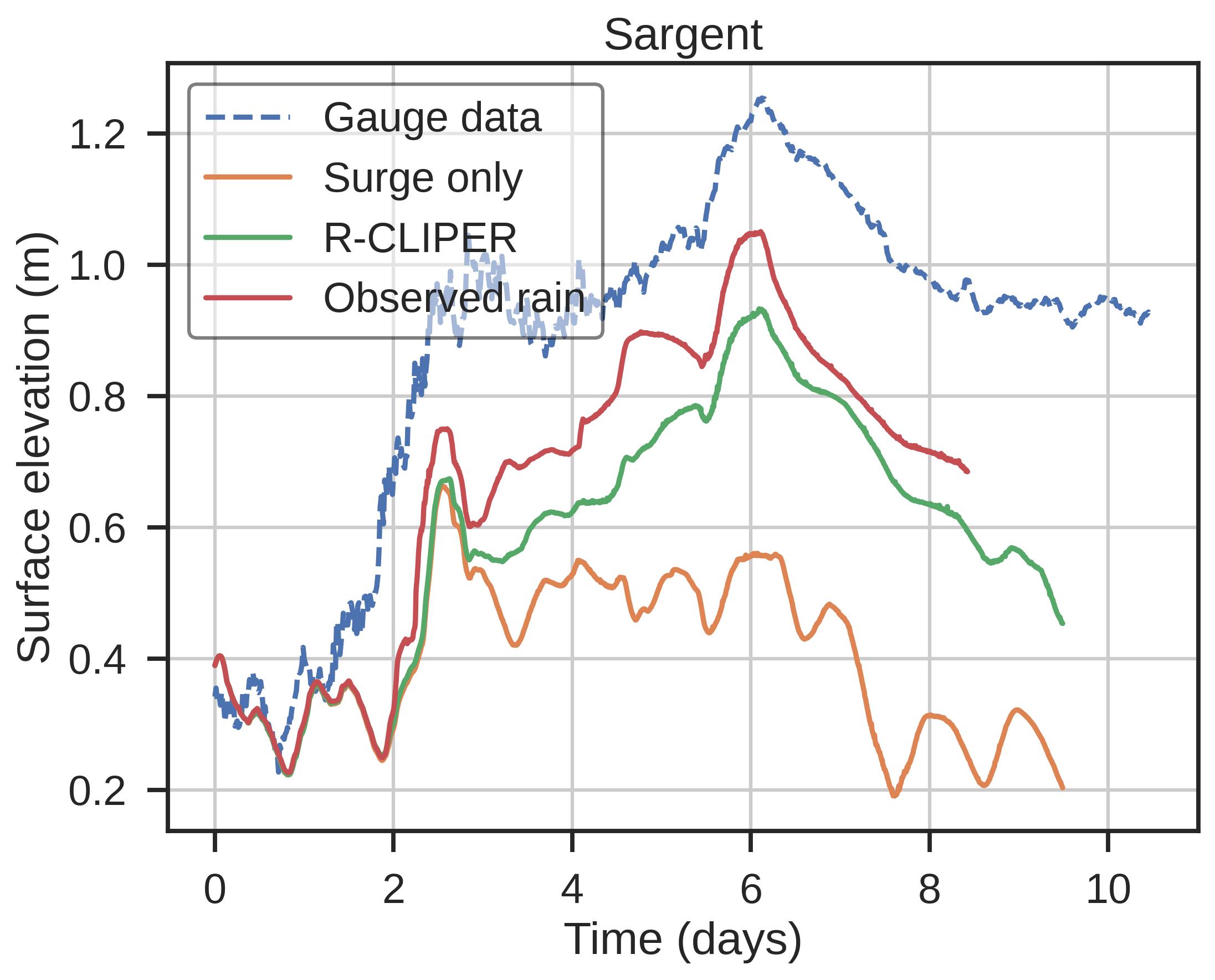}
    \includegraphics[width=0.49\textwidth]{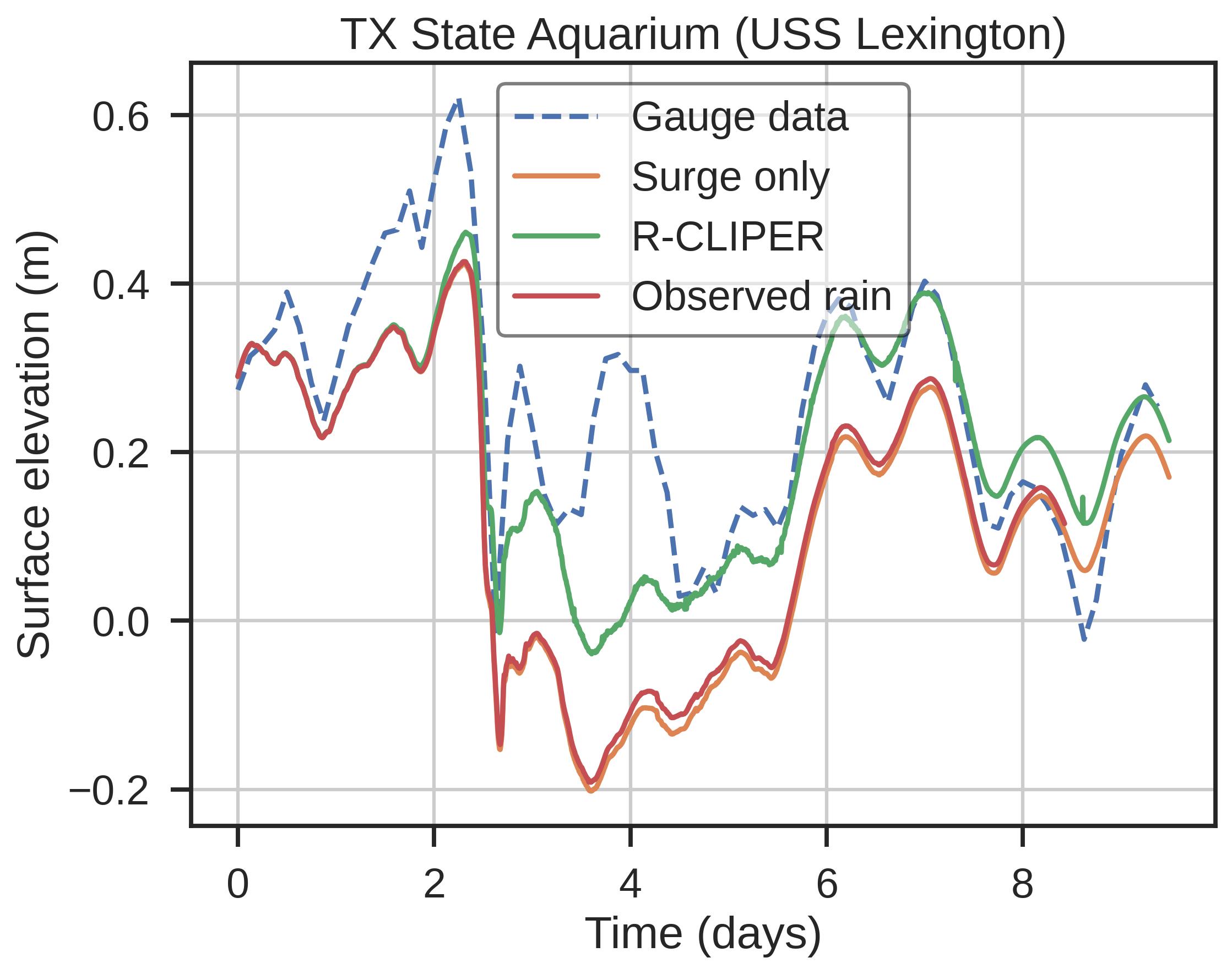}\\
    \includegraphics[width=0.49\textwidth]{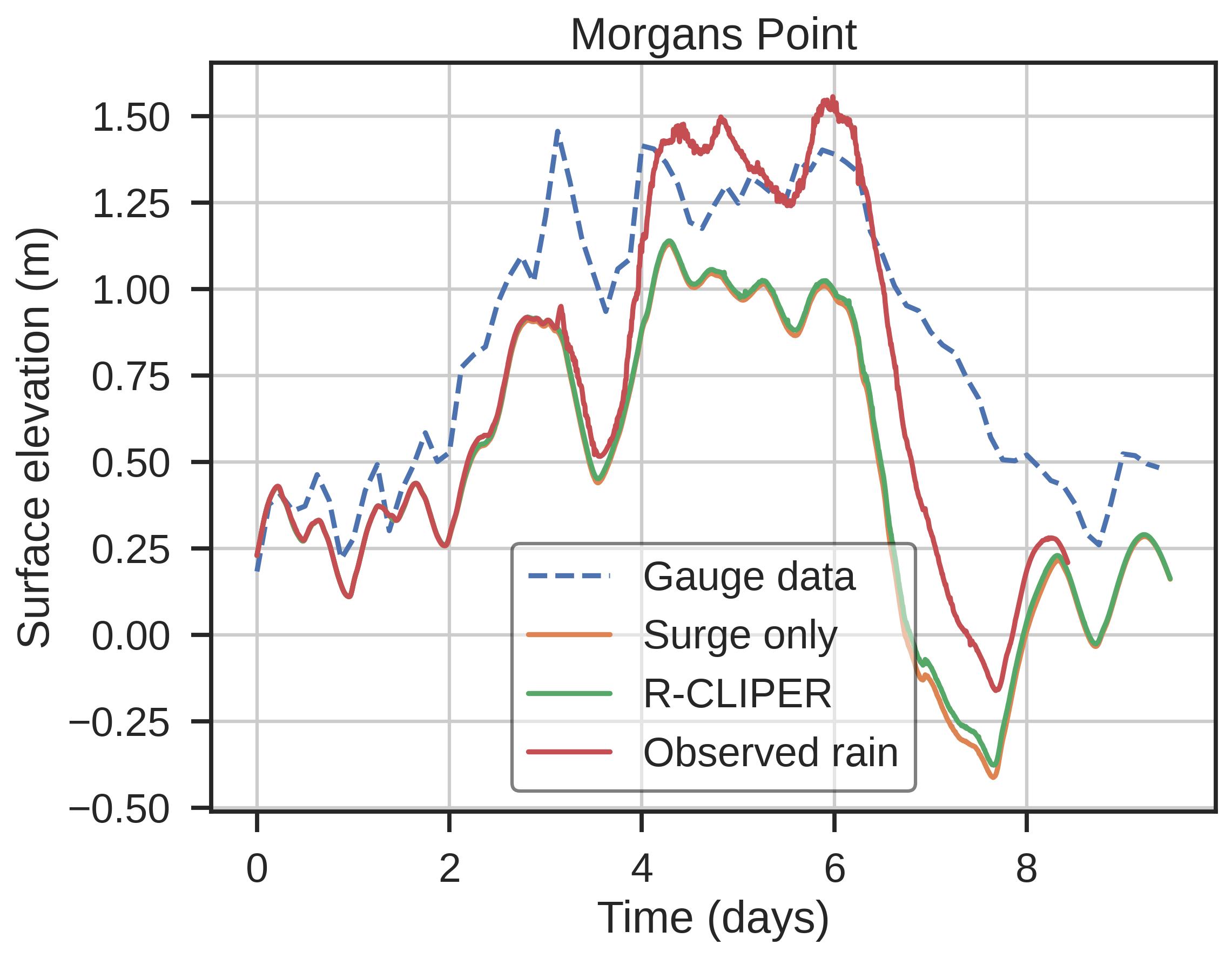}
    \includegraphics[width=0.49\textwidth]{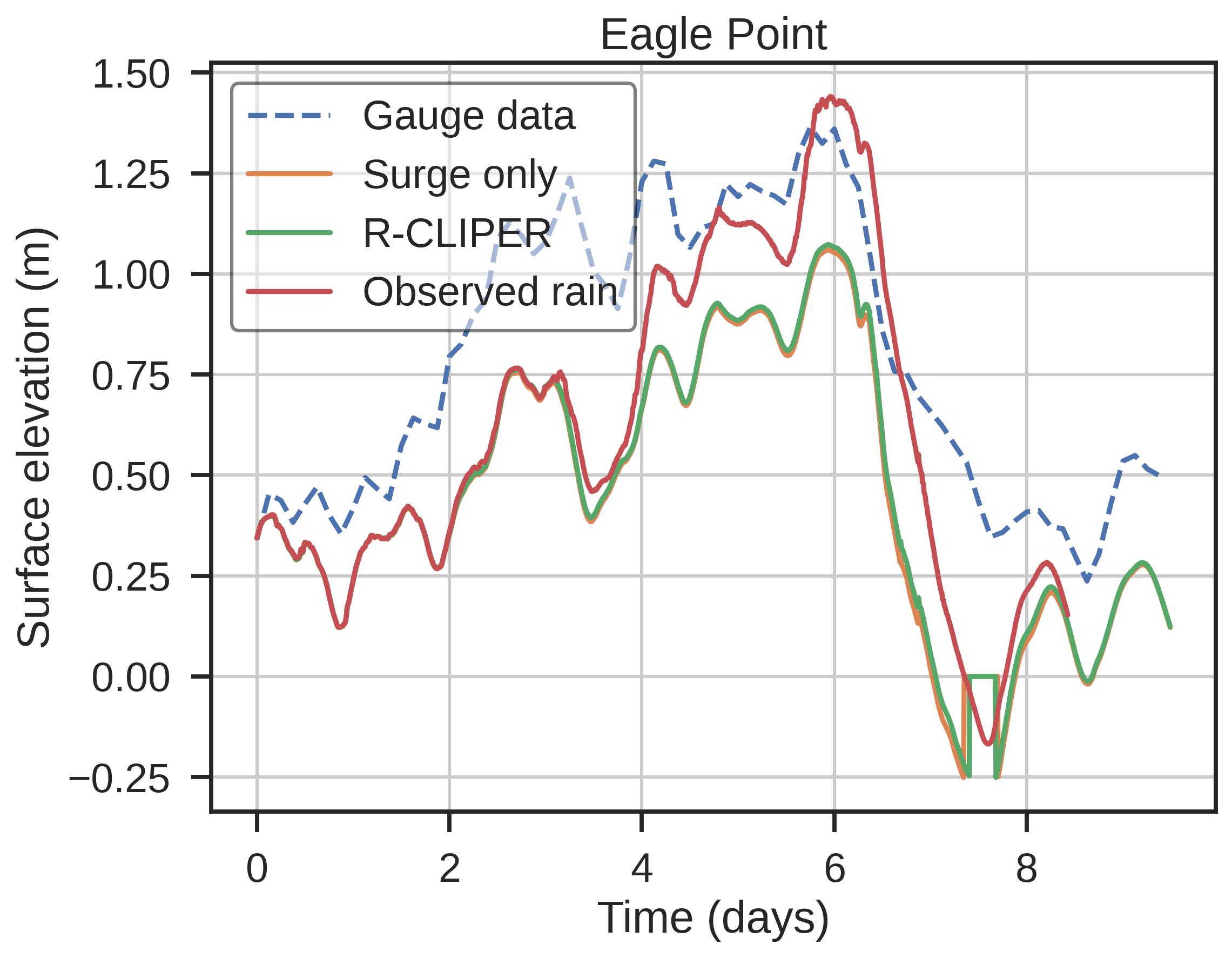}

    \caption{Water elevation comparisons between a run without any rain (orange), one using parametric rainfall (green), one using observed rain data (red), and NOAA station measurements (dashed blue). Time is relative to the starting date 8/23/2017.}
    \label{fig:elev}
\end{figure}
Figure \ref{fig:elev} presents a comparison between our model output and the water elevation at these stations. The corresponding root mean squared errors (RMSE) are shown in Table \ref{rmse}.
At all locations, we notice an underprediction of the water surface elevation in the surge-only case. The addition of rainfall improves the results by increasing the elevation at all considered locations and reducing their RMSE, although the discrepancy between R-CLIPER and observed rainfall appears to vary by location. In particular, at
Morgans Point and Eagle Point which are close to where the majority of the compound flood occurred, the R-CLIPER model made nearly no difference compared to the surge only case.
The reason can be deduced by inspecting its formulation in Figure~\ref{fig:RCLIPER} and the track of Hurricane Harvey in Figure~\ref{fig:stations}; since the eye of the hurricane passed very far off the coast, the contribution from R-CLIPER in the Houston area is vanishingly small.

\begin{figure}[h!]
    \centering
    \includegraphics[width=0.95\textwidth]{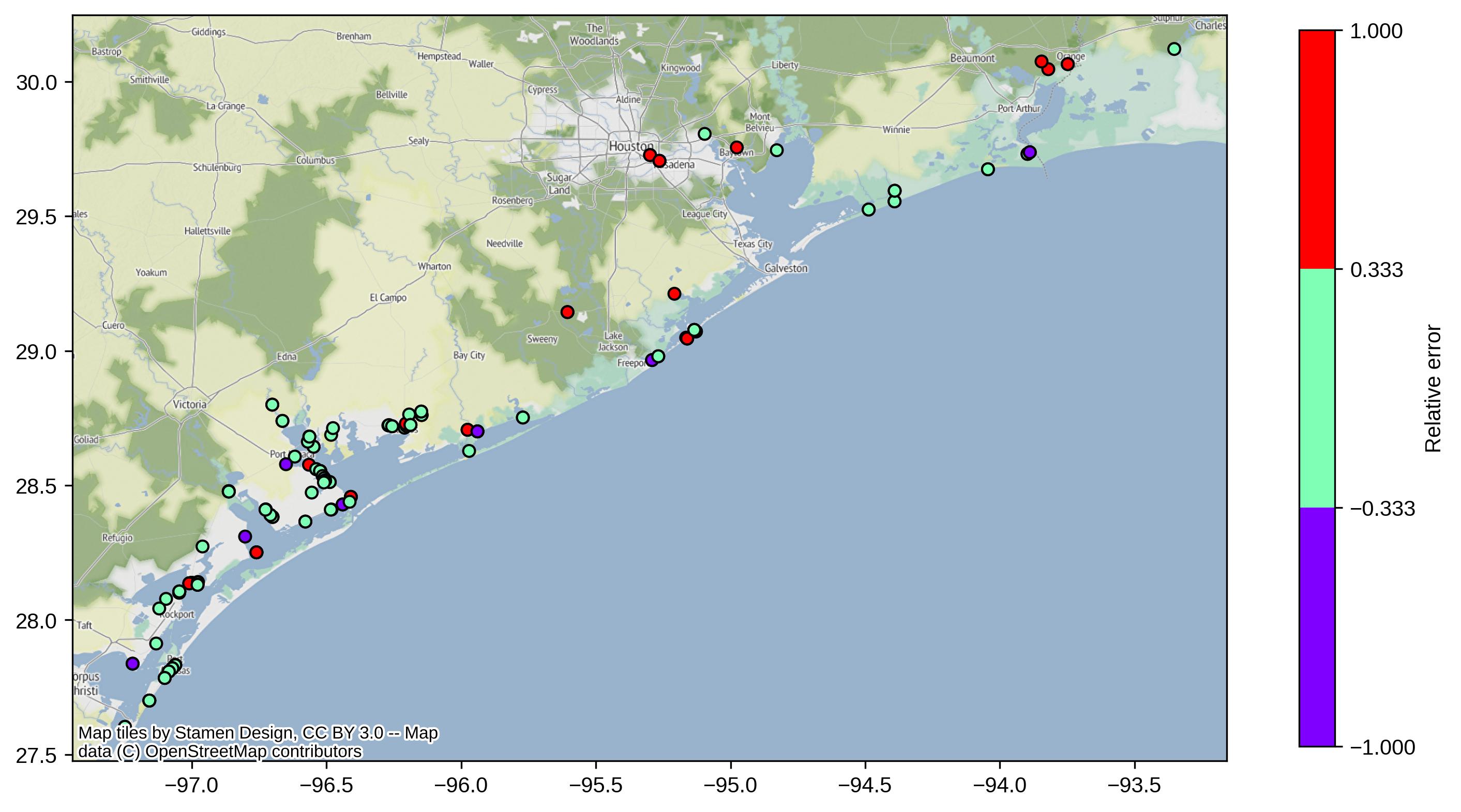} \\
    \includegraphics[width=0.95\textwidth]{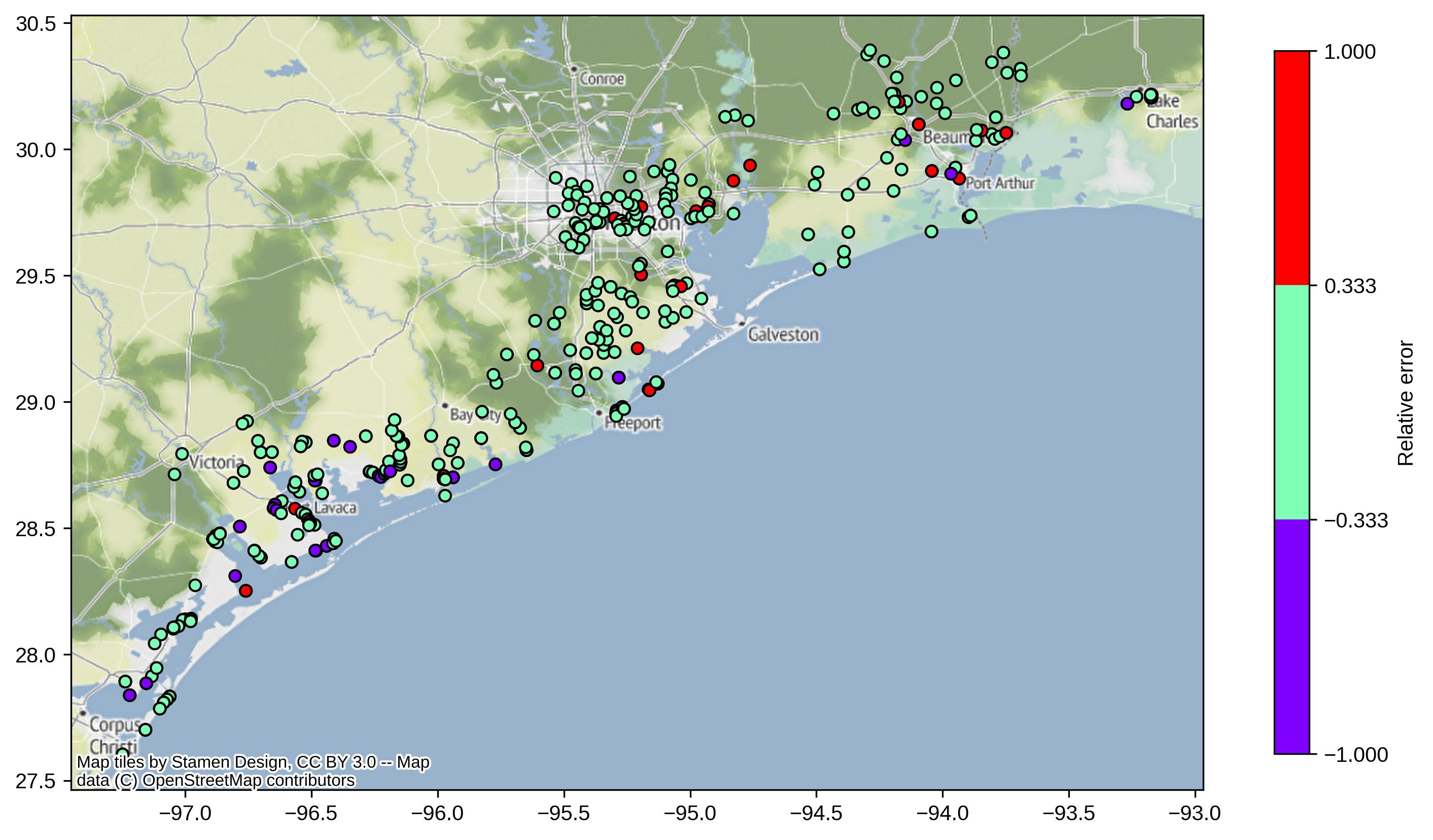} %

    \caption{\textbf{Top}: relative error between observed HWMs and output from DG-SWEM for the surge-only Hurricane Harvey case.
    \textbf{Bottom}: same plot but for the surge + R-CLIPER case. Only locations flagged as wet by DG-SWEM are shown. Positive relative error indicates underprediction while a negative value indicates overprediction. }
    \label{fig:hwm_err}
\end{figure}

\begin{figure}[h!]
    \centering
    \includegraphics[width=0.49\textwidth]{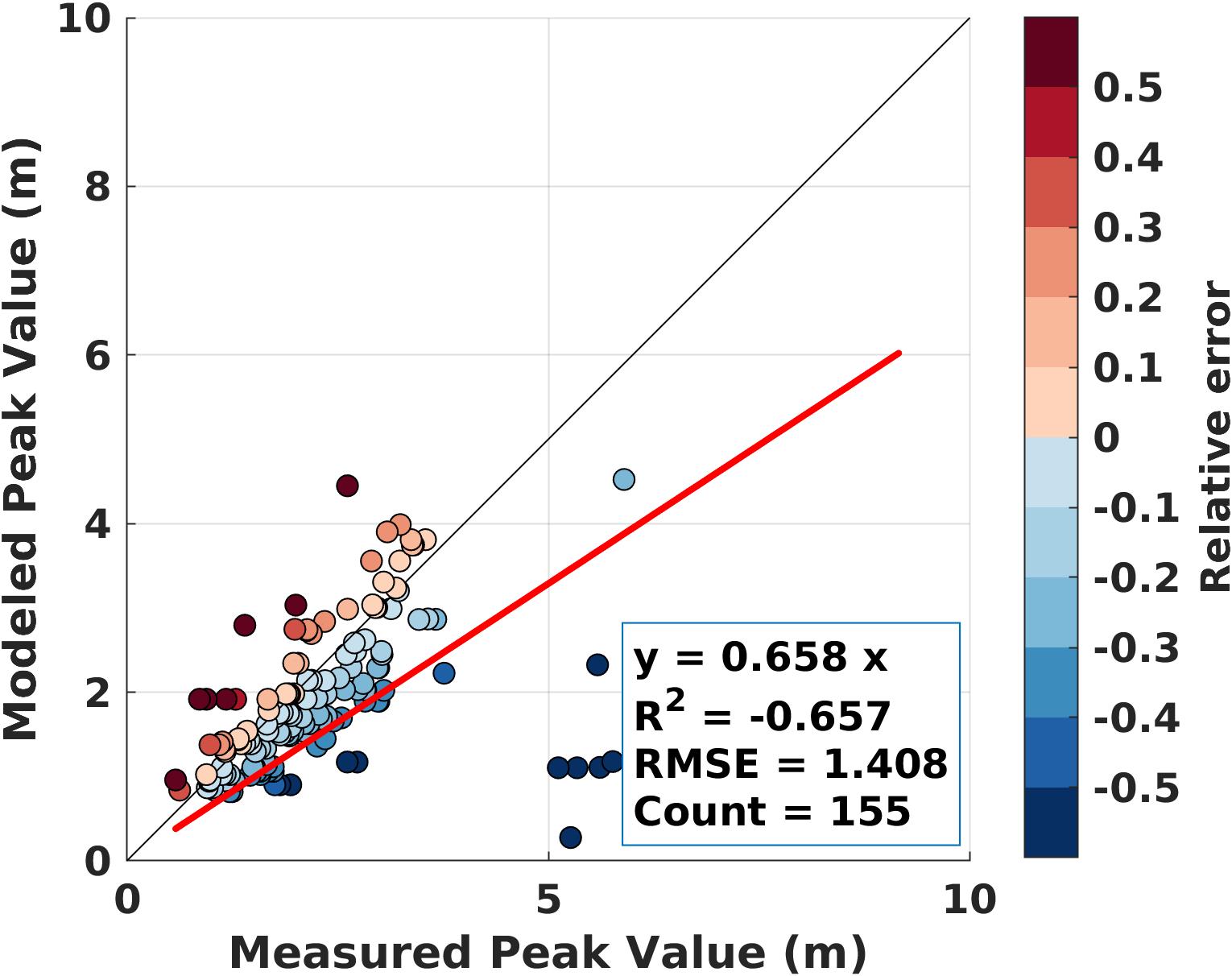}
    \includegraphics[width=0.49\textwidth]{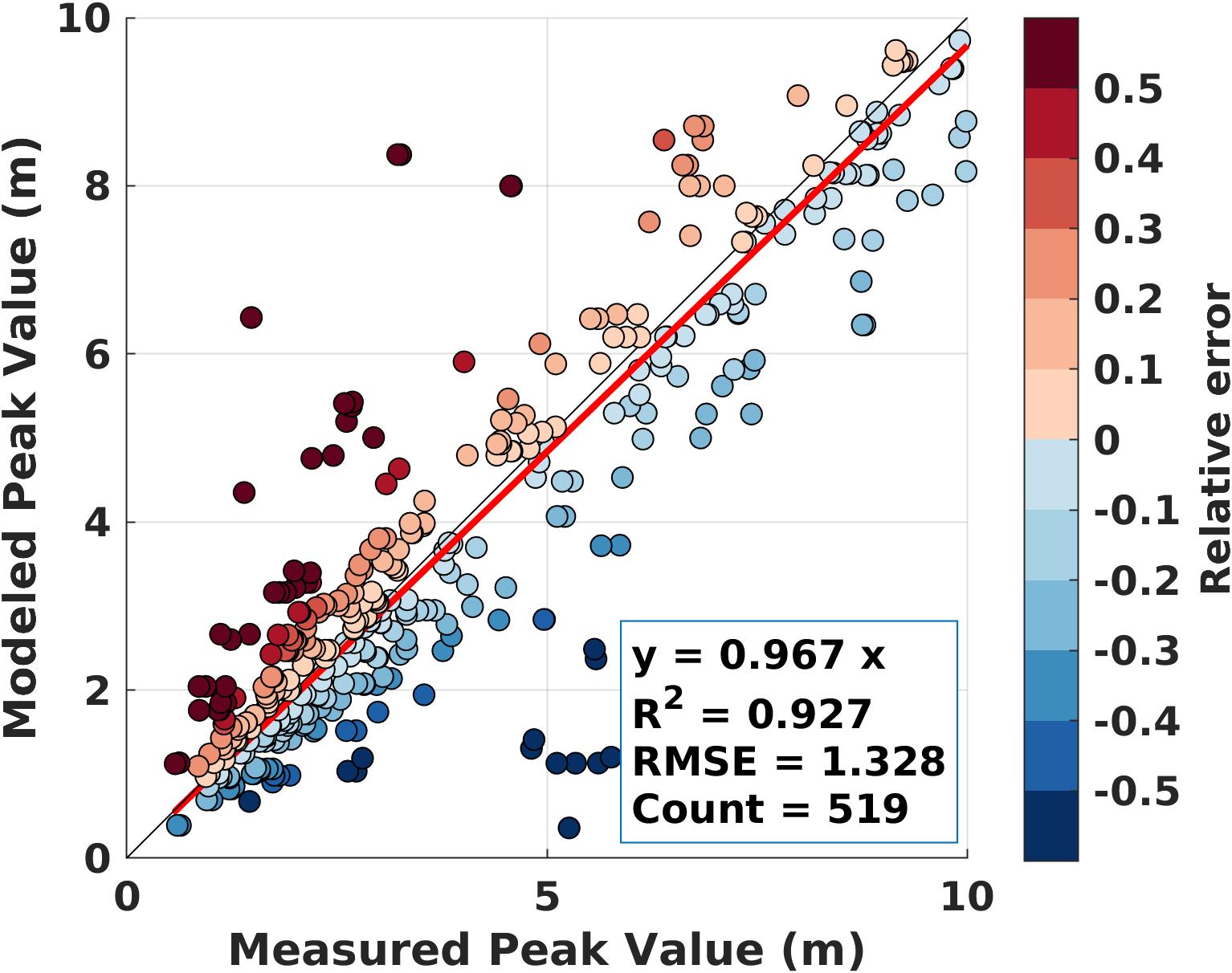} %
    \caption{High water mark comparison for Hurricane Harvey. \textbf{Left}: the surge-only case. \textbf{Right}: surge + R-CLIPER parametric rainfall model. Only wet locations from the model are included. }
    \label{fig:rain}
\end{figure}

\begin{table}
\centering
\begin{tabularx}{0.9\textwidth}{ X X X X }
    \hline
    & Surge only  & Surge + R-CLIPER & Surge + observed rain \\
    \hline
    Port Aransas & 0.156 & 0.119 & 0.157 \\
    Aransas Wildlife & 0.395 & 0.207 & 0.297 \\
    Sargent & 0.529 & 0.309 & 0.221 \\
    USS Lexington & 0.184 & 0.108 & 0.186 \\
    Morgans Point & 0.478 & 0.461 & 0.322 \\
    Eagle Point & 0.378 & 0.368 & 0.292 \\
    \hline
\end{tabularx}
\caption{RMSE  (m) of the time series at different stations for each scenario.}
\label{rmse}
\end{table}
To evaluate the overall performance of using R-CLIPER in our model, we compare the model output from scenarios 1 and 2 to the high water mark (HWM) measurements obtained from USGS \cite{hwm}. These measurements are filtered by their qualities (i.e. with uncertainty of up to 0.1 ft); the model values at these locations are obtained by interpolating from nodal maximum elevation. Note that only points from the model considered ``wet", i.e. the interpolated surface elevation results in positive depth, are compared.
The relative error of these HWMs for these two scenarios are shown in Figure \ref{fig:hwm_err} to indicate the locations of under- and overpredictions. Each point is computed using the formula $(y - \hat{y})/|y|$ where $y$ is the measured HWM and $\hat{y}$ the model output. In the surge-only case, most of the overpredictions reside on the coast while a significant number of underpredictions reside deeper on the mainland. In the surge + R-CLIPER case, both quantities increase in those areas, but we also see many more accurate measurements arising from the addition of rainfall. In particular, the Houston and Beaumont areas now see an increase in water levels in contrast to the surge-only case where they were mostly dry.
We also quantitatively compare $R^2$ and RMSE in Figure \ref{fig:rain} by plotting measured HWMs against the peak model outputs. The $R^2$ value noticeably increased from 0.6566 to 0.9273 with the addition of parametric rain, indicating that the compounding effect is described reasonably well by our model. The slight decrease in RMSE from 1.4081 to 1.3284 also suggests that, overall, the incorporation of rainfall does not exacerbate existing model errors.

A limitation of our model for this particular experiment can be observed by comparing these results with previous studies on Harvey. In \cite{valle2020compound}, the Regional Ocean Modeling System (ROMS) is used to simulate the Galveston Bay and Houston area by also incorporating river discharge from Buffalo Bayou and San Jacinto River. In particular, they demonstrate that at Morgans Point, much of the flood is due to river discharge (Figure 10). For this location their surge-only case results in around 1 m peak elevation which is similar to our results for scenarios 1 and 2. Once both rivers are included, they observe a peak closer to observation at around 1.5 m.
Another study \cite{Stephens_Timothy_A2022-qq} also simulates Harvey on a similar domain using the Adaptive Hydraulics Model (AdH).
They incorporate both rainfall and discharge from multiple rivers and obtained an $R^2$ of 0.99 and RMSE of 0.83 for their validation against HWMs in the area.
They also obtained more qualitatively accurate elevation time series at Morgans Point (NOAA 8771013) and Eagle Point (NOAA 8770613) \cite[Figure~6]{Stephens_Timothy_A2022-qq}.
These comparisons highlight the impact of river runoff on the compounding effect in the mainland area not captured by the parametric rain model alone.

%These results present two interesting features: $i)$ while including more points that become wet and severely flood, the incorporated rainfall also introduces more outliers in the data, $ii)$ the $R^2$ value is noticeable increased when including rainfall but the RMSE is also increased.
\clearpage

\begin{figure}[h!]
    \centering
    \includegraphics[width=0.8\textwidth]{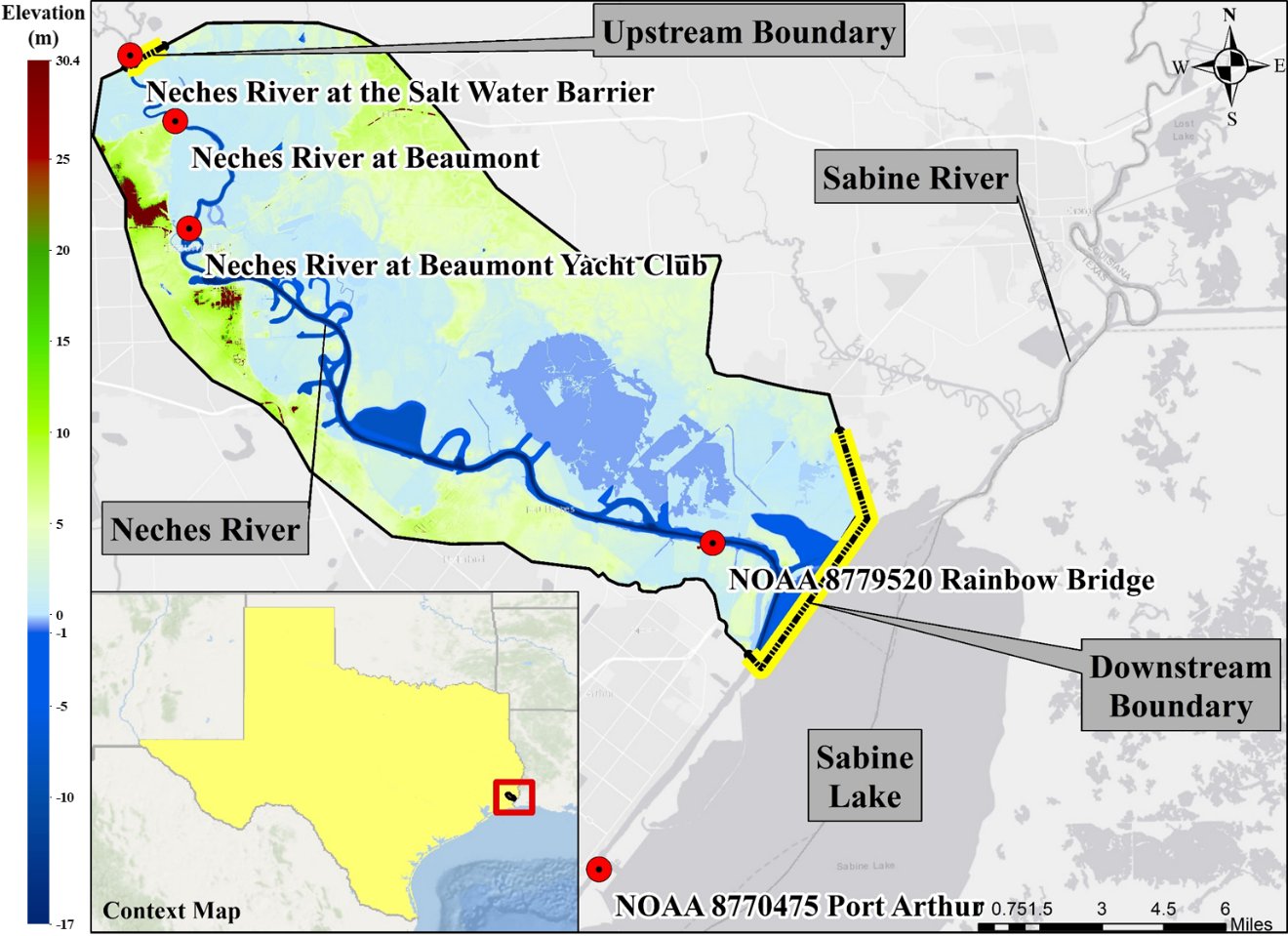}

    \caption{Computational domain of the Neches river case. Observation gauges are marked as red circles and boundaries are highlighted in yellow \cite{loveland2021developing}.}
    \label{fig:neches}
\end{figure}
\subsection{Neches test case}
The final test case is based on an approach to model compound flooding through river-coastal ocean interaction. In this case, we consider the Lower Neches River watershed  near the Texas-Louisiana border as shown in Figure~\ref{fig:neches} and highlighted in Figure~\ref{fig:harvey_track}. During Hurricane Harvey, this watershed was subjected to extreme rainfall and large portions of its surrounding areas were inundated during and after the event, see e.g.,~\cite{watson2018characterization} for details on the flooding during this event.
The model domain has been extracted from the  ADCIRC mesh used in the Harvey case and contains 122,839 elements and 62,075 nodes and is shown in Figure~\ref{fig:neches}.
Hence, the highly detailed unstructured mesh, bathymetry and topography, as well as parameters such as Manning's $n$  are preserved.
This event was also studied using ADCIRC and compared to HEC-RAS in~\cite{loveland2021developing} (see Section Methods - Study Area, Model Domains, and Inputs), and the present study uses the same input data as in that work. We only present key features of this model and refer readers to the original publication for greater detail.
\noindent As in \cite{loveland2021developing}, river flow data was extracted from a validated HEC-RAS model and a USGS gauge, and implemented as a flux boundary condition at the upstream. At the downstream end of the river which terminates into Sabine Lake, a time-varying elevation boundary condition was applied based on the same HEC-RAS model and the closest NOAA gauge. The locations of these inputs and the gauges used for validation are shown in Figure \ref{fig:neches}.

\begin{figure}[h!]
    \centering

    \includegraphics[width=0.49\textwidth]{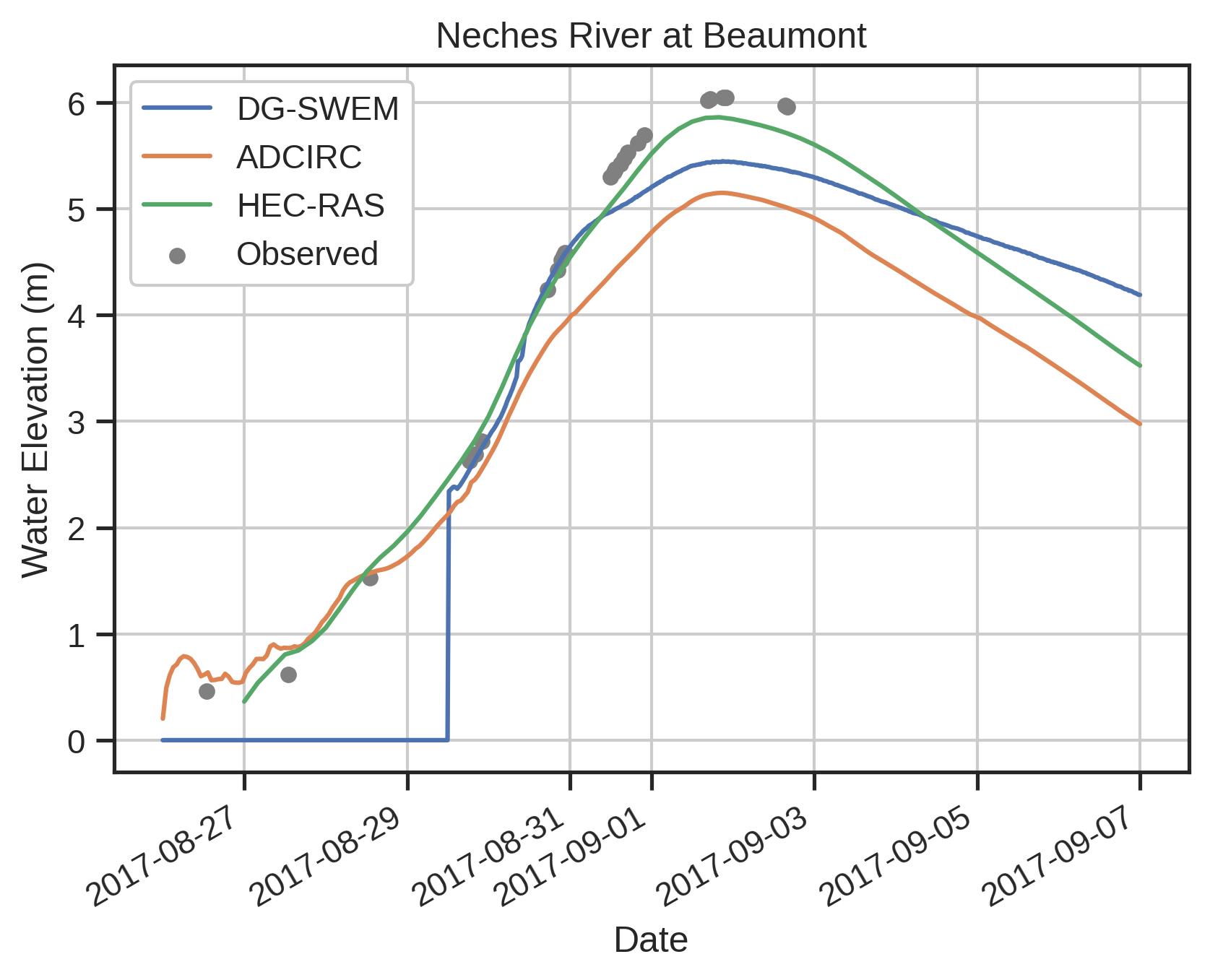}
  \includegraphics[width=0.49\textwidth]{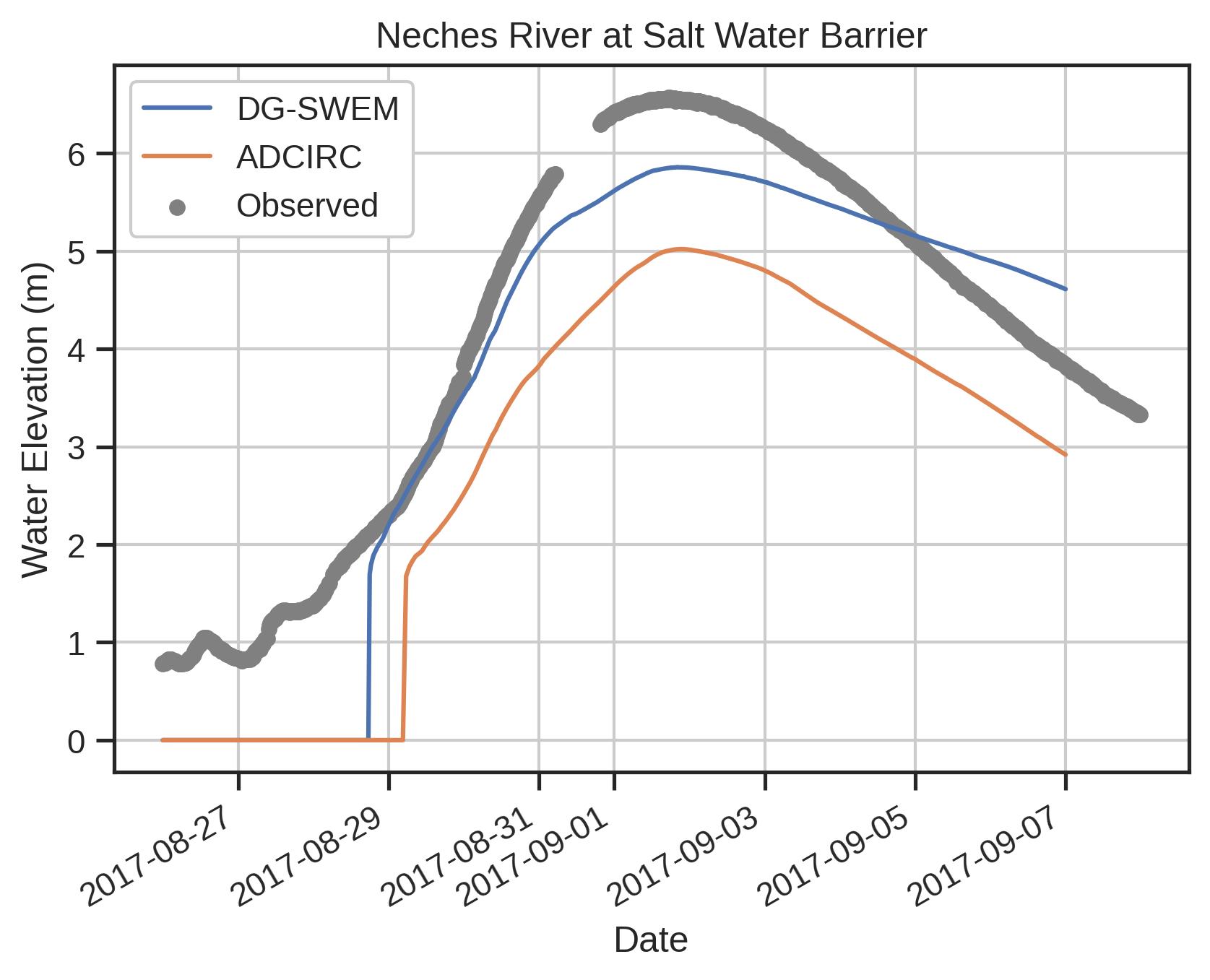}\\
      \includegraphics[width=0.49\textwidth]{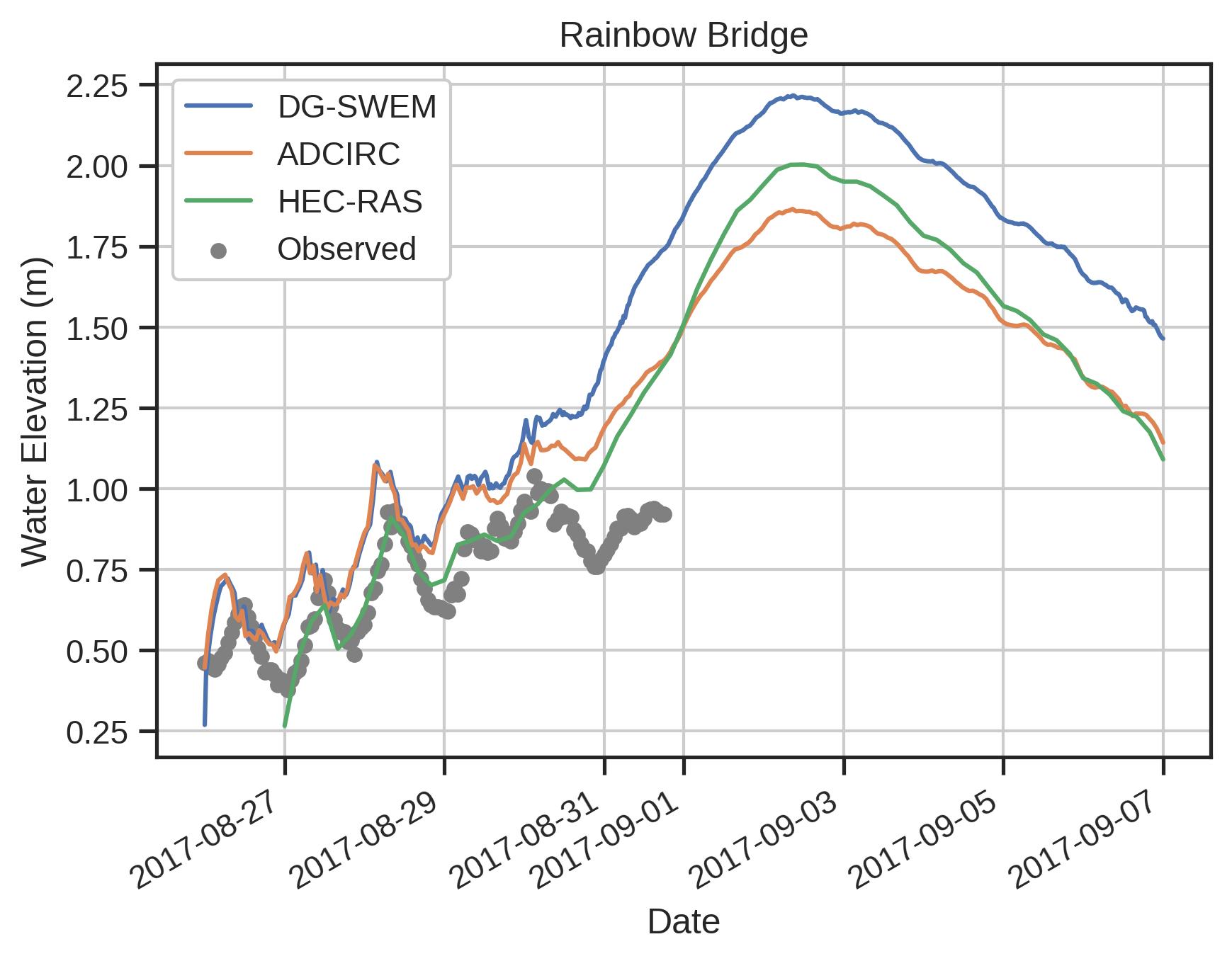}
    \includegraphics[width=0.49\textwidth]{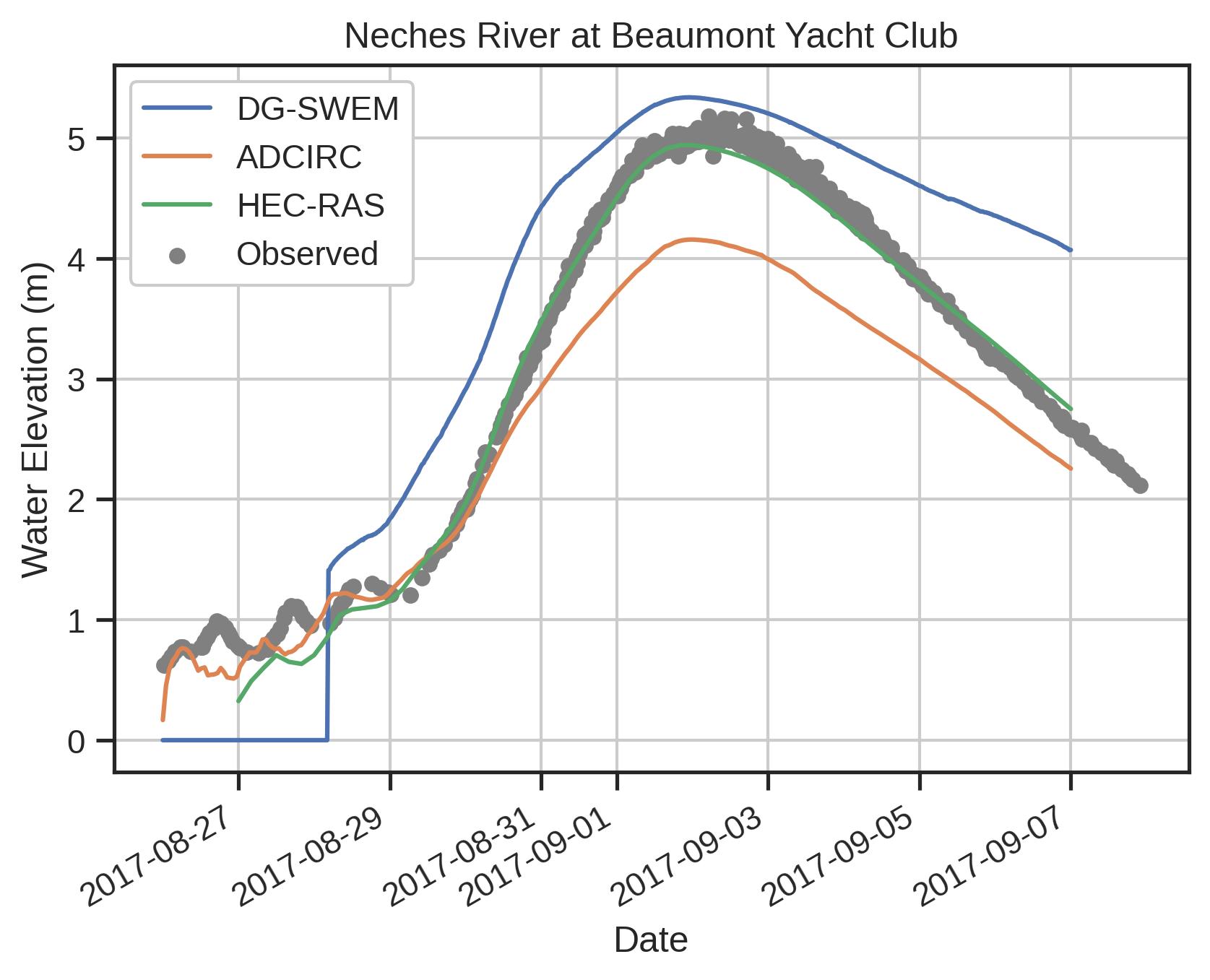}

    \caption{Simulated water elevation from DG-SWEM, ADCIRC and HEC-RAS as well as observed values at various stations along the Neches River. Flat portions shown with zero elevation indicate dry nodes, not necessary having zero elevation. Datum is NAVD88.}
    \label{fig:neches_gauges}
\end{figure}

Comparisons of outputs from DG-SWEM, ADCIRC, HEC-RAS (except at the Salt Water Barrier) as well as observed data from August 20 to September 1, 2017 are shown in Figure \ref{fig:neches_gauges}. We observe underprediction from ADCIRC in all cases except at the Rainbow Bridge. We also observe that in those cases, the output from DG-SWEM is higher and closer to peak observations. At Rainbow Bridge, DG-SWEM and ADCIRC match closely (while slightly overpredicting) up to the end of observed data.
This observation that DG-SWEM can more accurately handle highly advective flow has also been discussed in another study \cite{dawson2011discontinuous}. One explanation is that unlike DG-SWEM, ADCIRC solves a non-conservative reformulation of the continuity equation in (\ref{eq:SWE}) called the \textit{Generalized Wave Continuity Equation} (GWCE). This was done to avoid spurious oscillations associated with the continuous Galerkin formulation of the SWE \cite{Lynch1979-ux}, and thus we do expect differences in the output.
Similar to the previous study \cite{loveland2021developing}, the deviations of ADCIRC and DG-SWEM from HEC-RAS and observed data could be attributed to the fact that 1) the mesh used for ADCIRC and DG-SWEM are coarser than the HEC-RAS mesh at the stations and 2) the parameters in the HEC-RAS model (such as nodal attributes) have been specifically calibrated for this flow event, which is not the case for ADCIRC and DG-SWEM.
Finally, the mismatch between wet/dry elevation values at the beginning of DG-SWEM and ADCIRC can be attributed to differences in wetting-and-drying criteria as described in Section~\ref{sec:wetdry}.

\section{Concluding Remarks} \label{sec:conclusions}

In this paper, we have presented recent developments to compound flood modeling using DG methods. In particular, we exploit the conservation properties of DG methods to add rainfall as a source to the continuity equation in the shallow water equations. To ascertain spatially and temporally varying rainfall intensity we use parametric rainfall models from literature as well as interpolated rain data from past events.

We have shown results from extensive numerical experiments which highlight the capabilities and properties of our methodology, including conservation properties and compound flooding during a hurricane with significant rainfall.  In particular, we note the enhancements due to the addition of rainfall in the results for Hurricane Harvey (2017) in the areas close to the hurricane track, indicating the potential of using such parametric rainfall models in compound flood simulations. Comparisons to results from ADCIRC for river runoff in the Neches river further highlights  the capabilities of our DG methodology and the solution of the SWE.

While the DG methodology leads to accurate solutions, the increased number of degrees of freedom in the finite element approximation leads to significantly increased execution time when compared to e.g., ADCIRC. Thus, to make DG-SWEM more viable, we are currently incorporating GPU parallelization to the code as it has been shown to work well with the inherently local structure of the DG method \cite{Klockner2009-is, Fuhry2013-xg}. Future research directions of interest includes the combination of the DG method with Bubnov-Galerkin, e.g.,~\cite{dawson2002discontinuous} for increased efficiency as well as splitting methods for implicit-explicit time stepping. In~\cite{Tuleya:2007}, the authors also propose potential enhancements to the R-CLIPER model that will be investigated in future works to increase the fidelity of the simulations. The presented validation experiments focus on the comparison of time series and high water marks, other metrics, see e.g.,~\cite{clark2021abuse} could also be considered when investigating uncertain hydrologic processes.

Lastly, we note that while the addition of the rainfall and resulting runoff to this numerical DG model is a significant step towards modeling compound floods, there are a plethora of other hydraulic and hydrological processes that may also impact compound flood events  not explicitly accounted for here. These include, e.g., evapotranspiration, infiltration, and interception. Inclusion of these will be the focus of future works on the further extension of our model.

\section*{Acknowledgements}
This work has been supported by the United States National Science Foundation NSF PREEVENTS Track 2 Program, under NSF Grant Numbers. 1855047 (PI: Dawson) and 1854991 (PI: Kubatko).
This material is also based on work supported by the US Department of Homeland Security under Grant No.2015-ST-061-ND0001-01. The views and conclusions contained in this document are those of the authors and should not be interpreted as necessarily representing the official policies, either expressed or implied, of the US Department of Homeland Security.

The authors also would like to gratefully acknowledge the use of the ``ADCIRC", ``DMS23001", and ``DMS21031" allocations on the Frontera supercomputer at the Texas Advanced Computing Center at the University of Texas at Austin. The authors also would like to thank Dr. Mark Loveland for sharing some of the data used to create Figures \ref{fig:neches} and \ref{fig:neches_gauges}.

\bibliographystyle{elsarticle-num}
%\bibliography{references}


\end{document}